\documentstyle[12pt,a4wide,epsf]{article}
\begin{document}

\newcommand{\lsim}{\mbox{\raisebox{-.9ex}{~$\stackrel{\mbox{$<$}}{\sim}$~}}}
\newcommand{\gsim}{\mbox{\raisebox{-.9ex}{~$\stackrel{\mbox{$>$}}{\sim}$~}}}

\begin{center}
{\Large\boldmath\bf The Curvaton Hypothesis and the $\eta$-problem
of Quintessential Inflation, with and without Branes}

\bigskip

{\large\sc Konstantinos Dimopoulos$^{1,2}$}

\bigskip

{\em $^1$Physics Department, Lancaster University, Lancaster LA1 4YB, U.K.\\
$^2$Department of Physics, University of Oxford, Kebble Road, Oxford OX1 3RH, 
U.K.}

\bigskip

{\bf Abstract}
\end{center}

\noindent
{\small It is argued why, contrary to expectations, steep brane-inflation 
cannot really help in overcoming the $\eta$-problem of quintessential 
inflation model-building. In contrast it is shown that the problem is 
substantially ameliorated under the curvaton hypothesis. This is quantified by 
considering possible modular quintessential inflationary models in the context 
of both standard and brane cosmology.}

\section{Introduction}

Recent high redshift SN-Ia observations suggest that the Universe at present
is undergoing accelerated expansion \cite{snia}. These findings are consistent 
with the latest precise observations on the anisotropy of the Cosmic Microwave
Background Radiation (CMBR) \cite {cmb} and also with the observations of the 
Large Scale Structure (LSS) distribution of galactic clusters and 
superclusters \cite{conc}. 
Consequently, modern cosmology seems to have reached a point of concordance, 
which may be characterized by the following: We seem to live on a spatially
flat, homogeneous and isotropic Universe which, at present, is comprised by
about 1/3 of pressureless matter (dark matter mostly) and 2/3 by some other 
substance, with negative pressure, referred to as dark energy. The nature of 
this dark energy, however, remains elusive.

The above picture is in excellent agreement with the inflationary paradigm, 
which was initially introduced to solve the horizon and flatness problems 
of the Standard Hot Big Bang (SHBB) (and some other problems that were thought 
to be important at the time, such as the monopole problem) 
\cite{inf}\cite{book}. Inflation, in general, predicts a spatially 
flat Universe and also provides a superhorizon spectrum of curvature 
perturbations that result in adiabatic density perturbations which can 
successfully seed the formation of the observed LSS and the CMBR anisotropy. 
The spectrum of the curvature perturbations is predicted to be very near scale 
invariance, which agrees remarkably with the latest data. Hence, the 
inflationary paradigm is now considered by most cosmologists as the necessary 
extension of the SHBB, in order to form the Standard Model of Cosmology.

The successes of the inflationary paradigm have motivated many authors to
consider a similar type of solution to the dark energy problem at present.
Thus, it has been suggested that the current accelerated expansion of the
Universe is due to a late-time inflationary period driven by the potential
density of a scalar field $Q$, called quintessence (the fifth element, added 
to cold dark matter, hot dark matter (neutrinos), baryons and photons)
\cite{Q}. The aim for introducing quintessence was to avoid resurrecting the 
embarrassing issue of the cosmological constant $\Lambda$, which, if called 
upon to account for the dark energy, would have to be fine-tuned to the 
incredible level of \mbox{$\Lambda^2\sim 10^{-120}M_P^4$}, where $M_P$
is the Planck mass, i.e. the natural scale for Einstein's $\Lambda$.

However, it was soon realized that quintessence suffered from its own 
fine-tuning problems \cite{KL}. Indeed, in fairly general grounds it can be 
shown that at present \mbox{$Q\sim M_P$} (if originally at zero) with a mass 
\mbox{$m_Q\sim 10^{-33}$eV}, which is very hard to understand in the context 
of supergravity theories, where we expect the flatness of the potential to
be lifted on internal-space distances of the order of $M_P$. In addition, 
the introduction of yet again another unobserved scalar field (on top of the
inflaton field which drives the early Universe inflationary period) seems
unappealing. Finally, a rolling scalar field introduces another tuning 
problem, namely that of its initial conditions.

A compelling way to overcome the difficulties of the quintessence scenario is
to link it with the rather successful inflationary paradigm. This is quite 
natural since both inflation and quintessence are based on the same idea; that
the Universe undergoes accelerated expansion when dominated by the potential 
density of a scalar field, which rolls down its almost flat potential. This
unified approach has been named quintessential inflation \cite{PV} and is 
attained by identifying $Q$ with the inflaton field $\phi$. In quintessential 
inflation the scalar potential of $\phi$ is such that it causes two phases of 
accelerated expansion, one at early and the other at late times. 

However, the task of formulating such a potential is not easy and this is why 
not many successful attempts exist in the literature 
\cite{PV}\cite{ng}\cite{qmodels}\cite{jose}\cite{mine2}\cite{ipl}\cite{2exp}. 
Indeed, successful quintessential inflation has to account not only for the 
requirements of both inflation and quintessence \cite{yahiro} but also for a 
number of additional considerations. In particular, the minimum of the 
potential (taken to be zero, otherwise there is no advantage over the 
cosmological constant alternative) must not have been reached yet by the 
rolling scalar field, in order for the residual potential density not to be 
zero at present. This requirement is typically satisfied by potentials, which 
have their minimum displaced at infinity, 
\mbox{$V(\phi\rightarrow\infty)\rightarrow 0$}, a feature referred to as 
``quintessential tail''. Thus, quintessential inflation is a non-oscillatory 
inflationary model \cite{NO}. Another requirement is that of a ``sterile'' 
inflaton, whose couplings to the Standard Model (SM) particles are strongly 
suppressed. This is necessary in order to ensure the survival of the inflaton 
until today, so that it can become quintessence. Thus, in quintessential 
inflation the inflaton field does not decay at the end of the inflationary 
period into a thermal bath of SM particles. Instead, the reheating 
of the Universe is achieved through gravitational particle production during 
inflation, a process refereed to as gravitational reheating 
\cite{grreh}\cite{JP}. Because gravitational reheating can be 
rather inefficient, the Universe remains $\phi$-dominated after the end of
inflation, this time by the kinetic energy density of the scalar field. This 
period, called kination \cite{JP} (or deflation \cite{spok}), 
soon comes to an end and the Universe enters the radiation 
dominated period of the SHBB. Note, here that a sterile inflaton avoids the 
danger of violation of the equivalence principle at present, associated with 
coupled quintessence \cite{coupled}, where the ultra-light $Q$ corresponds to 
a long--range force. 

In the models \cite{PV}\cite{ng}\cite{qmodels} the plethora of constraints and 
requirements which are to be satisfied by quintessential inflation is managed 
through the introduction of a multi-branch scalar potential, that is a 
potential that changes form while the field moves from the inflationary to the 
quintessential part of its evolution. This change is either fixed ``by hand'' 
(such as in the original model \cite{PV}) or it is due to a potential with
different terms that dominate each at a time \cite{ng} or it is an outcome of 
a phase transition, arranged through some interaction of the inflaton with 
some other scalar fields \cite{qmodels}. Clearly this requires the 
introduction of a number of mass scales and couplings, which have to be tuned 
accordingly to achieve the desired results. Thus, in such models it is 
difficult to dispense with the fine-tuning problems of quintessence. Attempts 
to design a single-branch potential in \cite{jose}, which incorporates 
natural-sized mass scales and couplings have provided with existence proofs, 
but the the class of potentials presented are rather complicated. This is due 
to the so--called $\eta$-problem of quintessential inflation: Namely the fact 
that it is almost impossible to formulate a successful quintessential 
inflationary model with an inflationary scale high enough to satisfy the 
requirements of Big Bang Nucleosynthesis (BBN) but which neither results in 
strong deviations from scale invariance in the curvature perturbations 
spectrum, nor does it need to go over to super-Planckian
inflationary scale to solve the horizon problem. The $\eta$-problem is due to
the fact that between the inflationary plateau and the quintessential tail 
there is a difference of over a hundred orders of magnitude. To prepare for 
such an abysmal ``dive'' the scalar potential cannot help being strongly 
curved near the end of inflation, which destroys the scale invariance of the 
curvature perturbations. 

It has been thought that this problem is alleviated when considering inflation
in the context of brane-cosmology. Indeed, brane-cosmology allows for 
overdamped steep inflation \cite{steep}, which dispenses with the need for an 
inflationary plateau and, therefore, a curved potential seems no longer 
necessary. However, attempts to use this idea have still encountered 
difficulties (see for example \cite{mine2}\cite{ipl}) and the most promising 
results were achieved again with a multi-branch potential (a sum of
exponential terms) \cite{2exp}. In this paper we explain why. It seems that, 
despite the advantages of steep inflation, brane cosmology back-reacts by 
creating problems in the kination period. Indeed, we will show that the 
overdamping effect due to the modified dynamics of the Universe, inhibits the 
efficiency of kination in achieving a small late-time potential density. 

Fortunately, there is another solution to the $\eta$-problem of quintessential
inflation. Indeed, we show that the $\eta$-problem is substantially ameliorated
when considering inflation in the context of the curvaton hypothesis 
\cite{curv}. As shown recently in \cite{mine}, the curvaton hypothesis 
liberates inflationary models from the strains of the so-called {\sc cobe} 
constraint, i.e. the requirement that the amplification of the inflaton's 
quantum fluctuations during inflation should generate a curvature perturbation 
spectrum with amplitude that matches the observations of the Cosmic Background 
Explorer ({\sc cobe}). The curvaton hypothesis attributes the generation of 
the curvature perturbations to another scalar field, called the curvaton,
changing, thus, the {\sc cobe} constraint into an upper bound. In \cite{mine} 
it has been shown that this effect is rather beneficial
to many models of inflation well motivated by particle physics. Here, we 
demonstrate that it may assist also quintessential inflation in overcoming
the $\eta$-problem. This is because, in the context of the curvaton hypothesis, 
a curved potential does not necessarily destroy the scale-invariance of the 
curvature perturbation spectrum. Moreover, it may allow for significant 
reduction of the inflationary scale, which also proves beneficial for 
quintessential inflation.

The paper is organized as follows. In Section~2 the dynamics of the Universe
is briefly layed out both in the case of conventional and also brane cosmology.
In Section~3 we look in more detail into the period of kination, which is 
crucial for quintessential inflation. In Section~4 we discuss the motivation, 
characteristics and merits of the exponential quintessential tail, which we 
adopt throughout the paper. In Section~5 we describe the $\eta$-problem and
demonstrate that brane-cosmology cannot overcome it because it inhibits 
kination. In order to show this we calculate the constraints imposed on
quintessential inflation by the BBN and coincidence requirements. We also
study the constraints due to the possible overproduction of gravity waves.
In Section~6 we present the alternative idea in order to overcome the
$\eta$-problem, namely the curvaton hypothesis. In Section~7 we demonstrate
the curvaton liberating effects on a variant of modular inflation in the 
context of conventional cosmology. We calculate in detail the allowed 
parameter space and show that all the relevant requirements are met. In 
Section~8 we investigate the curvaton liberation effects in the case of 
brane-cosmology, using an exponential potential. We find that successful 
quintessential inflation is possible in a certain range of values for the 
brane tension. We carefully calculate the allowed parameter space and show how 
all the requirements and constraints are satisfied. Finally, in Section~9 we 
discuss our results and present our conclusions. Throughout the paper we use 
units such that \mbox{$c=\hbar=1$} in which Newton's gravitational constant is 
\mbox{$G=M_P^{-2}$}, where \mbox{$M_P=1.22\times 10^{19}$GeV}.

\section{Dynamics with and without Branes}

To set the stage for quintessential inflation let us briefly discuss the 
dynamics of the Universe in both conventional and brane cosmology.

The Universe is usually modeled as a collection of perfect fluids.
The background fluid, with density \mbox{$\rho_B\equiv\rho_\gamma+\rho_m$}
is comprised of relativistic matter (or radiation), with density $\rho_\gamma$
and pressure \mbox{$p_\gamma=\frac{1}{3}\rho_\gamma$}, and non-relativistic 
matter (or just matter), with density $\rho_m$ and pressure $p_m=0$. In 
addition we will consider a homogeneous scalar field $\phi$, which can be 
treated as a perfect fluid with density 
\mbox{$\rho_\phi\equiv\rho_{\rm kin}+V$} and pressure 
\mbox{$p_\phi\equiv\rho_{\rm kin}-V$}, where \mbox{$V=V(\phi)$} is the 
potential density and 
\mbox{$\rho_{\rm kin}\equiv\frac{1}{2}\dot{\phi}^2$} is the kinetic density of
$\phi$ respectively, with the dot denoting derivative with respect to the 
cosmic time $t$.

For every component of the Universe content one defines the barotropic 
parameter as \mbox{$w_i\equiv p_i/\rho_i$}. Energy momentum conservation
demands

\begin{equation}
d(a^3\rho)=-pd(a^3)
\label{conserve}
\end{equation}
which, for decoupled fluids, gives

\begin{equation}
\rho_i\propto a^{-3(1+w_i)}
\label{barotrop}
\end{equation}
where $a$ is the scale factor of the Universe. To study the dynamics of the 
Universe one also needs the equation of motion of the scalar field:

\begin{equation}
\ddot{\phi}+3H\dot{\phi}+V'=0
\label{field}
\end{equation}
where \mbox{$H\equiv\dot{a}/a$} is the Hubble parameter and
the prime denotes derivative with respect to $\phi$.

In standard cosmology the global geometry of the Universe is described by
the Friedman-Robertson-Walker (FRW) metric. The temporal component of
the Einstein equations for this metric is the Friedman equation:

\begin{equation}
H^2=\frac{\rho}{3m_P^3}
\label{fried}
\end{equation}
where \mbox{$m_P\equiv M_P/\sqrt{8\pi}$} is the reduced Planck mass and we have
considered a spatially flat Universe, according to observations. Using 
(\ref{conserve}) and (\ref{fried}) one obtains

\begin{eqnarray}
H=\frac{2t^{-1}}{3(1+w)} & \qquad
a\propto t^{\frac{2}{3(1+w)}} \qquad &
\rho=\frac{4}{3(1+w)^2}\left(\frac{m_P}{t}\right)^2
\label{evol}
\end{eqnarray}
where $w$ corresponds to the dominant component of the Universe content and, 
in the above, \mbox{$w\neq -1$}. In the case of cosmological constant 
domination \mbox{$\rho=m_P^2\Lambda=$ const.} so that (\ref{barotrop}) gives 
\mbox{$w=-1$}. Then (\ref{fried}) becomes 
\mbox{$H^2=\frac{1}{3}\Lambda=$ const.} and, therefore, 
\mbox{$a\propto\exp(Ht)$}, i.e. the Universe undergoes pure de-Sitter 
inflation.

The above dynamics is substantially modified if one considers a Universe with
at least one large extra dimension. In particular we will concern ourselves
with the, so--called, second Randall-Sundrum scenario, in which our Universe 
is a four-dimensional submanifold (brane) of a higher-dimensional space-time.
Matter fields are confined on this brane but gravity can propagate also in
the extra dimensions (bulk). The simplest realization of this scenario 
considers a five-dimensional space-time, i.e. one large extra dimension.
In this case standard cosmology can be recovered in low energies if one 
considers that the density and pressure on the brane are given by 
\mbox{$\rho_b\equiv\rho+\lambda$} and 
\mbox{$p_b\equiv\rho-\lambda$} respectively, i.e. the brane is endowed with
a constant tension $\lambda$ \cite{cline0}. 
The brane tension $\lambda$ is related to the 
fundamental (5-dimensional) Planck mass $M_5$ by

\begin{equation} 
\lambda=\frac{3}{4\pi}\left(\frac{M_5^3}{M_P}\right)^2
\label{M5}
\end{equation}
Then the analogous to the Friedman equation is \cite{binedefa}:

\begin{equation}
H^2=\frac{1}{3}\Lambda+\frac{\rho}{3m_P^2}\left(1+\frac{\rho}{2\lambda}\right)
+\frac{\cal E}{a^4}
\label{friedbrane}
\end{equation}
where $\cal E$ is a constant of integration, related to bulk gravitational 
waves or black holes in the vicinity of the brane (dark radiation), which is 
usually inflated away during the first few e-foldings of brane-inflation. The 
4-dimensional cosmological constant $\Lambda$ is due to both the brane tension 
$\lambda$ and the (negative) bulk cosmological constant $\Lambda_5$. $\Lambda$ 
can be tuned to zero by demanding \mbox{$\lambda=-\Lambda_5m_P^2$}. Similarly 
to conventional thinking this $\Lambda$ tuning is considered to be due to some 
unknown symmetry. In view of the above we can recast (\ref{friedbrane}) as

\begin{equation}
H^2=\frac{\rho}{3m_P^2}\left(1+\frac{\rho}{2\lambda}\right)
\label{Hfric}
\end{equation}
which reduces to the usual Friedman equation (\ref{fried}) when 
\mbox{$\rho\ll\lambda$}, so that standard cosmology is recovered. 
However, for energies \mbox{$\rho\gg\lambda$} the above becomes

\begin{equation}
H=\frac{\rho}{\sqrt{6\lambda}\,m_P}
\label{bfried}
\end{equation}

As a result of the above, the dynamics of the Universe is modified for energy 
higher than the brane tension. Since the matter fields are confined on the
brane, energy conservation for matter and radiation on the brane is retained 
and (\ref{conserve}) is still valid. Then, in view of (\ref{bfried}), we 
obtain 

\begin{eqnarray}
H=\frac{t^{-1}}{3(1+w)} & \qquad
a\propto t^{\frac{1}{3(1+w)}} \qquad &
\rho=\frac{\sqrt{6\lambda}}{3(1+w)}\left(\frac{m_P}{t}\right)
\label{bevol}
\end{eqnarray}
Thus, we see that the effect of the extra dimension is to reduce the
rate of Hubble expansion for energy larger than $\lambda$. This will also
affect the evolution of the scalar field $\phi$ because (\ref{field}) shows
that $H$ generates a friction term for the roll-down of the field.

To complete our discussion for the Universe dynamics we need to mention that
the temperature of the Universe is, at any time, given by

\begin{equation}
\rho_\gamma=\frac{\pi^2}{30}g(T)T^4
\label{rgT}
\end{equation}
where $g(T)$ is the number of relativistic degrees of freedom that corresponds 
to the thermal bath of temperature $T$. At high temperatures 
\mbox{$g\sim 10^{-2}$}, whereas at present \mbox{$g_0=3.36$}.

\section{Kination}\label{kin}

Kination is a period of the Universe evolution, when $\rho$ is dominated by
the kinetic density of the scalar field \cite{JP}\cite{spok}. Kination is one 
of the essential ingredients of quintessential 
inflation because it allows the field to rapidly roll down its potential,
reducing its potential density substantially, so that the huge gap between
the inflationary energy density and the density at present is possible to 
bridge. In order for kination to occur it is necessary that the reheating 
process is not prompt. Fortunately, this is exactly what we expect when 
considering a sterile inflaton.

\subsection{After the end of inflation}

\subsubsection{Gravitational reheating}

Since a sterile inflaton field does not decay at the end of inflation, 
after the inflationary period most of the energy density of the Universe is
still in the inflaton. The thermal bath of the Standard Hot Big Bang 
(SHBB) is due to the gravitational production of particles during inflation.
This process is known as gravitational reheating \cite{grreh}, and results in  
density \mbox{$\rho_{\rm reh}\sim 10^{-2}H_{\rm end}$}, where `end' denotes 
the end of inflation. The gravitationally produced particles soon therlmalize 
so that, in view of (\ref{rgT}), we can define a reheating temperature 
$T_{\rm reh}$ such that

\begin{equation}
\rho_{\rm reh}=\frac{\pi^2}{30}g_{\rm reh}T_{\rm reh}^4
\label{rreh}
\end{equation}
where $g_{\rm reh}\sim 10^2$ is the number of relativistic degrees of
freedom at reheating. In the Standard Model (SM) \mbox{$g_{\rm reh}=106.75$}. 
However, in supersymmetric extensions of the SM $g_{\rm reh}$ is at least twice 
as large (e.g. in the MSSM \mbox{$g_{\rm reh}=229$}). 

The gravitational reheating temperature is determined by the Gibbons-Hawking 
temperature in de Sitter space, which gives 

\begin{equation}
T_{\rm reh}\equiv\alpha\left(\frac{H_{\rm end}}{2\pi}\right)
\label{Treh}
\end{equation}
where $\alpha$ is the reheating efficiency. For purely gravitational reheating
\mbox{$\alpha\sim 0.1$}. However, even tiny couplings of the inflaton with 
another field may increase $\alpha$ dramatically \cite{prehqinf} and can even 
lead to parametric resonance effects (instant preheating 
\cite{instant}\cite{NO}), which would result in
\mbox{$\alpha\gg 1$}. The reheating efficiency will prove crucial to our
considerations, so we will retain it as a free parameter, since it is, in 
principle, determined by the underlying physics of the quintessential
inflationary model.

In the above we implicitly assumed that the thermalization of the
gravitationally produced particles is instantaneous. This is not really so,
which means that the actual reheating temperature may be somewhat (about an 
order of magnitude) smaller than the estimate of (\ref{Treh}). However, this
will not really affect our treatment because the scaling of 
\mbox{$\rho_B\simeq\rho_\gamma$} after the end of inflation does not have to do
with whether $\rho_\gamma$ is thermalized or not.

\subsubsection{The onset of the Hot Big Bang}

Gravitational reheating is typically a very inefficient process so that
\mbox{$\rho_\phi^{\rm end}\gg\rho_B^{\rm end}$}. However, because at the end 
of inflation \mbox{$V_{\rm end}\simeq\rho_{\rm kin}^{\rm end}$}, the inflaton 
soon becomes dominated by its kinetic density 
\mbox{$\rho_\phi\simeq\rho_{\rm kin}\gg V$}, which means that 
\mbox{$w_\phi\approx 1$} and the Universe is characterized by 
a stiff equation of state. In this case (\ref{conserve}) suggests that 
\mbox{$\rho\propto a^{-6}$}. In contrast, 
\mbox{$\rho_B\simeq\rho_\gamma\propto a^{-4}$}, which means that eventually 
the density of the background thermal bath will come to dominate the Universe.
At this time the SHBB begins. 

Note that, when it is kinetic density dominated, the scalar field becomes 
entirely oblivious of its potential density as its field equation 
(\ref{field}) is dominated by the kinetic terms

\begin{equation}
\ddot{\phi}+3H\dot{\phi}\simeq 0
\label{fieldkin}
\end{equation}
This means that the scalar field evolution engages into a free-fall behaviour, 
which enables us below to study kination in a model independent way.

\subsection{Brane kination}

Let us first consider kination in the context of brane cosmology. We assume,
therefore that the inflationary density scale is larger than the brane tension,
i.e. \mbox{$V_{\rm end}\gg\lambda$}. In this case the Friedman equation is 
given by (\ref{bfried}) and the Universe evolves according to (\ref{bevol}).
Then the end of inflation takes place when 
\mbox{$\rho(t_{\rm end})=V_{\rm end}$}, which gives

\begin{equation}
t_{\rm end}=\sqrt{\frac{\lambda}{6}}\,\frac{m_P}{V_{\rm end}}
\label{btend}
\end{equation}

Using (\ref{bevol}) we find that, during kination, \mbox{$a\propto t^{1/6}$}.
Therefore, because after inflation \mbox{$\rho=\rho_{\rm kin}\propto a^{-6}$},
we find \mbox{$\dot{\phi}=\sqrt{2\,t_{\rm end}V_{\rm end}}\,t^{-1/2}$}, which
results in

\begin{equation}
\phi(t)=\phi_{\rm end}+
\frac{4}{\sqrt{6}}\left(\frac{\lambda}{2V_{\rm end}}\right)^{1/2}
\left(\sqrt{\frac{t}{t_{\rm end}}}-1\right)m_P
\label{bft1}
\end{equation}
where, without loss of generality, we assumed that \mbox{$\dot{\phi}>0$}.

According to (\ref{evol}) and (\ref{bevol}) the switch-over to conventional
cosmology occurs when \mbox{$\rho=\rho_\lambda$}, where

\begin{equation}
\rho_\lambda=\frac{1}{2}\,\lambda
\label{rl}
\end{equation}
which takes place at the time

\begin{equation}
t_\lambda=\frac{2m_P}{\sqrt{6\lambda}}
\label{tl}
\end{equation}

At this time, (\ref{bft1}) suggests that the field has rolled to the value 
%\mbox{$\phi_\lambda\equiv\phi(t_\lambda)$} given by

\begin{equation}
\phi_\lambda=\phi_{\rm end}+\frac{4}{\sqrt{6}}
\left(1-\sqrt{\frac{\lambda}{2V_{\rm end}}}\right)m_P
\label{fl}
\end{equation}

After $t_\lambda$ the Universe evolves according to the standard FRW cosmology.
Thus, using \mbox{$w=1$}, we find from (\ref{evol}) that 
\mbox{$a\propto t^{1/3}$}. Therefore, \mbox{$\rho\propto a^{-6}$} gives
\mbox{$\dot{\phi}=\frac{2}{\sqrt{6}}\,m_P/t$}. Hence, for \mbox{$t>t_\lambda$} 
we find

\begin{equation}
\phi(t)=\phi_{\rm end}+\frac{2}{\sqrt{6}}\left[
2\left(1-\sqrt{\frac{\lambda}{2V_{\rm end}}}\right)+
\ln\left(\frac{t}{t_\lambda}\right)\right]m_P
\label{bft2}
\end{equation}

The end of kination occurs when 
\mbox{$\rho_{\rm kin}=\rho_\gamma=\frac{1}{2}\rho_*\equiv\rho(t_*)$}. Using the
scaling laws for $\rho_{\rm kin}$ and for $\rho_\gamma$ it is easy to find that
\mbox{$t_*=\sqrt{t_{\rm end}t_\lambda}(V_{\rm end}/\rho_{\rm reh})^{2/3}$}, or,
equivalently, that

\begin{equation}
t_*=\frac{1}{\sqrt{3}}\frac{m_PV_{\rm end}}{\rho_{\rm reh}^{3/2}}
\label{t*rreh}
\end{equation}

Employing (\ref{rreh}) we can recast the above as

\begin{equation}
t_*=\frac{(24\pi)^3}{\sqrt{3}\,\alpha^6}
\left(\frac{30}{g_{\rm reh}}\right)^{3/2}
\frac{\lambda^3m_P^7}{V_{\rm end}^5}
\label{bt*}
\end{equation}

Inserting this into (\ref{bft2}) we find that, by the end of kination,
the field has rolled to %\mbox{$\phi_*\equiv\phi(t_*)$}, where

\begin{equation}
\phi_*=\phi_{\rm end}+\frac{2}{\sqrt{6}}\left[
3\ln\!\left(\!\frac{48\pi}{\alpha^2}\sqrt{\frac{30}{g_{\rm reh}}}\right)\!+\!
2\!\left(\!1\!-\!\sqrt{\frac{\lambda}{2V_{\rm end}}}\right)\!-\!
\frac{7}{2}\ln\!\left(\frac{2V_{\rm end}}{\lambda}\right)\!+\!
\frac{3}{2}\ln\!\left(\frac{m_P^4}{V_{\rm end}}\right)\right]m_P
\label{bf*}
\end{equation}

Using (\ref{evol}) and (\ref{rgT}) we find the temperature at the end of 
kination

\begin{equation}
T_*=\frac{\alpha^3}{2(12\pi)^2}\sqrt{\frac{g_{\rm reh}}{5}}
\left(\frac{g_{\rm reh}}{g_*}\right)^{1/4}
\frac{V_{\rm end}^{5/2}}{\lambda^{3/2}m_P^3}
\label{bT*}
\end{equation}
where $g_*$ is the number of relativistic degrees of freedom at the end of 
kination. This is the temperature at the onset of the SHBB and therefore it has
to be constrained by BBN considerations: \mbox{$T_*>T_{\sc bbn}$}, where 
\mbox{$T_{\sc bbn}\sim 1$ MeV}. However, here it should be pointed out that
$T_*$ is overestimated above because we have considered an instantaneous
transition between inflation and kination. In reality this transition takes 
some time so that $t_*$ is somewhat larger and, therefore, $T_*$ turns out 
about an order of magnitude smaller that the estimate of (\ref{bT*}) 
\cite{ipl}. This will be taken into account when we apply the BBN constraint 
below. Note that, typically, $T_*$ is hard to be much larger than 
$T_{\sc bbn}$ and, therefore, \mbox{$g_*=10.75$}, which corresponds to the 
number of relativistic degrees of freedom just before pair annihilation.

\subsection{Conventional kination}

Working in a similar manner we can study kination in conventional cosmology.
This time $t_{\rm end}$ is decided by (\ref{evol}) and is found to be

\begin{equation}
t_{\rm end}=\frac{m_P}{\sqrt{3V_{\rm end}}}
\label{tend}
\end{equation}

There is a subtlety here which is worth mentioning. The time $t_{\rm end}$
is {\em not} the actual cosmic time interval that corresponds to the duration
of inflation. In fact, $t_{\rm end}$ is the age the Universe would have been
at the end of inflation {\em were it always kinetic density dominated}. 
This means that $t_{\rm end}$ is always ``normalized'' according to the
evolution stage that the Universe enters after the end of kination. Note, 
however, that there is a difference here between the brane and conventional 
cases, namely the fact that, for the same cosmic time $t$ and the same $w$, 
the Hubble parameter in conventional cosmology, as given by (\ref{evol}), is 
double the size of the one in brane cosmology, given by (\ref{bevol}). This 
fact reflects itself in the ``normalization'' of $t_{\rm end}$ as we will 
discuss below.

Using (\ref{tend}) and the scaling of $\rho_{\rm kin}$ we find that, during
kination

\begin{equation}
\phi(t)=\phi_{\rm end}+\frac{2}{\sqrt{6}}\ln\left(\frac{t}{t_{\rm end}}\right)
\label{ft}
\end{equation}

Let us now estimate the time $t_*$ when kination ends. It turns out that 
(\ref{t*rreh}) is still valid in the conventional kination case. Using 
this we obtain

\begin{equation}
t_*=\frac{(12\pi)^3}{\sqrt{3}\,\alpha^6}
\left(\frac{30}{g_{\rm reh}}\right)^{3/2}
\frac{m_P^7}{V_{\rm end}^2}
\label{t*}
\end{equation}
which suggests that 

\begin{equation}
\phi_*=\phi_{\rm end}+\frac{2}{\sqrt{6}}\left[
3\ln\left(\frac{12\pi}{\alpha^2}\sqrt{\frac{30}{g_{\rm reh}}}\right)+
\frac{3}{2}\ln\left(\frac{m_P^4}{V_{\rm end}}\right)\right]m_P
\label{f*}
\end{equation}

Finally, the temperature at the end of kination is 

\begin{equation}
T_*=\frac{\alpha^3}{(12\pi)^2}\sqrt{\frac{2g_{\rm reh}}{5}}
\left(\frac{g_{\rm reh}}{g_*}\right)^{1/4}
\frac{V_{\rm end}}{m_P^3}
\label{T*}
\end{equation}

From the above one can see that the conventional kination results may be
obtained by the brane kination ones if we take 
\mbox{$\rho_\lambda\rightarrow V_{\rm end}$} 
(i.e. \mbox{$\lambda\rightarrow 2V_{\rm end}$}) and 
\mbox{$\alpha\rightarrow 2\alpha$}. the latter is due to the difference of the
``normalization'' of $t_{\rm end}$ between the brane and the conventional 
case, which has been discussed above.

\subsection{The Hot Big Bang}

After the end of kination the Universe becomes radiation dominated, but the
scalar field continues to be dominated by its kinetic density. Therefore,
\mbox{$\rho_\phi=\rho_{\rm kin}\propto a^{-6}$}, but now 
\mbox{$a\propto t^{1/2}$} according to (\ref{barotrop}) for 
\mbox{$w=w_\gamma=\frac{1}{3}$}. Using (\ref{fieldkin}) we find

\begin{equation}
\phi(t)=\phi_*+\frac{4}{\sqrt{6}}\left(1-\sqrt{\frac{t_*}{t}}\right)m_P
\end{equation}

Thus, in about a Hubble time the kinetic density of the scalar field 
is entirely depleted and the field freezes to the value

\begin{equation}
\phi_F=\phi_*+\frac{4}{\sqrt{6}}m_P
\end{equation}

Using (\ref{bf*}) and (\ref{f*}) we obtain $\phi_F$ in the brane and 
conventional cases respectively:

\begin{equation}
\phi_F=\phi_{\rm end}+\frac{2}{\sqrt{6}}\left[4+
3\ln\!\left(\!\frac{48\pi}{\alpha^2}\sqrt{\frac{30}{g_{\rm reh}}}\right)-
2\sqrt{\frac{\lambda}{2V_{\rm end}}}-\!
\frac{7}{2}\ln\!\left(\frac{2V_{\rm end}}{\lambda}\right)\!+\!
\frac{3}{2}\ln\!\left(\frac{m_P^4}{V_{\rm end}}\right)\right]m_P
\label{bfF}
\end{equation}

and 

\begin{equation}
\phi_F=\phi_{\rm end}+\frac{2}{\sqrt{6}}\left[2+
3\ln\left(\frac{12\pi}{\alpha^2}\sqrt{\frac{30}{g_{\rm reh}}}\right)+
\frac{3}{2}\ln\left(\frac{m_P^4}{V_{\rm end}}\right)\right]m_P
\label{fF}
\end{equation}

\section{The exponential quintessential tail}

It can be shown that a quintessential tail with a milder than exponential
slope results in eternal acceleration \cite{jose}. However, string theory 
disfavours eternal acceleration because it introduces future horizons which 
inhibit the determination of the S-matrix \cite{string}. Moreover, mild 
quintessential tails, e.g. of the inverse power-law type \cite{ipl}, make it 
hard to satisfy coincidence. Furthermore, quintessential tails steeper than 
exponential have disastrous attractors \cite{jose}, which, not only do they 
diminish $\rho_\phi$ faster than $\rho_B$, but are also reached very soon 
after the end of inflation and, therefore, cannot lead to late-time 
acceleration. According to the above, the best chance we have for achieving 
successful quintessential inflation is by considering potentials
with exponential quintessential tails of the form:

\begin{equation}
V(\phi\gg\phi_{\rm end})\simeq V_0\exp(-b\phi/m_P)
\label{exptail}
\end{equation}
where $b$ is a positive constant whose value is crucial to the behaviour of the
system. 

With the use of (\ref{field}) it can be shown that a potential of the above 
form has an attractor solution \mbox{$\phi(t)=\phi_{\rm attr}$} such that

\begin{equation}
\phi_{\rm attr}(t)=\frac{2}{b}\ln\left[
\sqrt{\frac{V_0}{2}\left(\frac{1+w}{1-w}\right)}\,\frac{bt}{m_P}\right]m_P
\label{fattr}
\end{equation}

The field follows the attractor solution after
\mbox{$\phi_{\rm attr}(t)=\phi_F$}, when it unfreezes and begins rolling in
accordance to (\ref{fattr}). The attractor solution results in
\mbox{$\rho_{\rm kin}^{\rm attr}=2b^{-2}(m_P/t)^2$} and

\begin{equation}
V_{\rm attr}(t)=\frac{2}{b^2}\left(\frac{1-w}{1+w}\right)
\left(\frac{m_P}{t}\right)^2
\label{Vattr}
\end{equation}

Thus, we see that 
\mbox{$\rho_\phi^{\rm attr}=\rho_{\rm kin}^{\rm attr}+V_{\rm attr}$} scales
in the same way as $\rho(t)$ as given by (\ref{evol}). Therefore, there are
two possible cases:

\medskip

\noindent {\bf i) Subdominant scalar.}
In this case \mbox{$w=w_B$} and \mbox{$\rho\propto a^{-3(1+w_B)}$}, which 
means that

\begin{equation}
\frac{\rho_\phi}{\rho_B}=\frac{3}{b^2}(1+w_B)
\end{equation}

Since \mbox{$\rho_\phi<\rho_B$} we see that

\begin{equation}
b^2>3(1+w_B)
\end{equation}

\noindent {\bf ii) Dominant scalar.}
In this case \mbox{$w=w_\phi$} and \mbox{$\rho\propto a^{-3(1+w_\phi)}$}. 
This time \mbox{$\rho(t)=\rho_\phi$}, which, 
in view of (\ref{evol}), gives

\begin{equation}
b^2=3(1+w_\phi)
\end{equation}

Now, the acceleration of the Universe expansion is determined by the spatial 
component of the Einstein equations, which, for the spatially flat FRW 
metric, is:

\begin{equation}
\frac{\ddot{a}}{a}=-\frac{\rho+3p}{6m_P^2}
\end{equation}

Therefore, the Universe will engage into accelerated expansion if 
\mbox{$\rho_\phi>\rho_B$} and \mbox{$w_\phi<-\frac{1}{3}$}. Thus, we can avoid
eternal acceleration, even in the case of the dominant scalar if 
\mbox{$b^2>2$}. Hence, dark energy domination without eternal acceleration 
can be achieved if $b$ lies in the range

\begin{equation}
2<b^2<3(1+w_B)
\label{brange0}
\end{equation}

\begin{figure}
\begin{center}
\leavevmode
\hbox{
\epsfxsize=5in
\epsffile{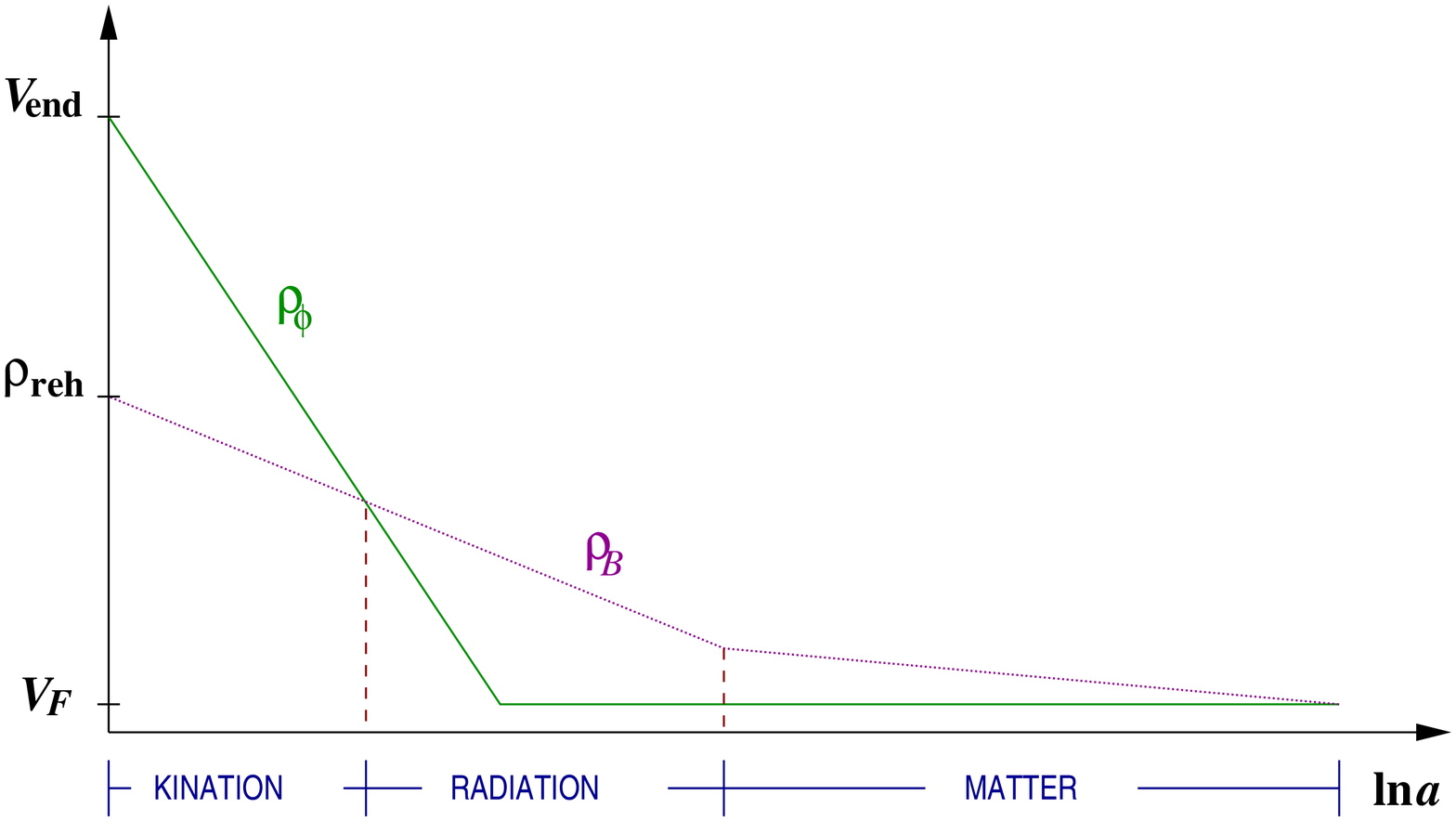}}
\end{center}
\caption{\footnotesize
The scaling of $\rho_\phi$ (solid line) and $\rho_B$ (dotted line) after the 
end of inflation. Initially, 
\mbox{$\rho_\phi=V_{\rm end}\gg\rho_{\rm reh}=\rho_B$} and we have kination. 
However, because \mbox{$\rho_\phi=\rho_{\rm kin}\propto a^{-6}$}, whereas
\mbox{$\rho_B=\rho_\gamma\propto a^{-4}$}, eventually $\rho_B$ dominates 
$\rho_\phi$ and kination ends. Afterwards $\rho_\phi$ continues to be 
$\rho_{\rm kin}$--dominated until all the kinetic energy is depleted and the 
field freezes at constant density \mbox{$\rho_\phi=V_F$}. In the meantime 
the radiation era continues and \mbox{$\rho_B=\rho_\gamma\propto a^{-4}$}.
Much later, though, matter density takes over and the radiation era ends.
In the matter era \mbox{$\rho_B=\rho_m\propto a^{-3}$}. The matter era
continues until today when \mbox{$\rho_B\simeq V_F\simeq\rho_0$} and the 
scalar field becomes important again.}
\label{kin1fig}
\end{figure}

However, it has been shown that even though eternal acceleration is avoided,
the Universe does accelerate for a brief period when the attractor is reached 
and the field unfreezes from $\phi_F$ to follow it. If the scalar field density
becomes dynamically important before it has begun following the attractor it is
still characterized by a barotropic parameter \mbox{$w_\phi\approx -1$} and,
therefore it will cause some acceleration. In fact, numerical simulations have
shown that this is possible even in the subdominant scalar case if the 
attractor is not too far below $\rho_B$, i.e. if $b$ is not too large. This is
due to the fact that the system, after unfreezing, oscillates briefly around
the attractor path in phase space, before settling down to follow it, as
have been shown by numerical simulations \cite{oscil}. This 
behaviour has been shown to enlarge the effective range of $b$, which may 
avoid eternal but achieve brief acceleration. According to \cite{cline} brief 
acceleration may be achieved if

\begin{equation}
2<b^2<24
\label{brange}
\end{equation}
Therefore, in order to explain the observed accelerated expansion of the 
Universe, the scalar field has to unfreeze at present, i.e. we require 
\mbox{$V_F\equiv V(\phi_F)$} to be

\begin{equation}
V_F=\Omega_\phi\rho_0
\label{coinc0}
\end{equation}
where \mbox{$\Omega_\phi\simeq 0.65$} is the observed fraction of the dark 
energy density over the present critical density $\rho_0$. The above is usually
called the `coincidence' constraint. The scaling of
$\rho_\phi$ and $\rho_B$ after the end of inflation is shown in 
Figure~\ref{kin1fig}.

\section{Coincidence versus BBN and Gravitational Waves}

We now investigate the requirement of successful coincidence in combination 
with the BBN constraint in both brane and conventional cosmology. These two
requirements are the most difficult to achieve in quintessential inflation.
This is because, one the one hand the BBN constraint pushes the inflationary 
scale towards high energies, while on the other hand the coincidence 
constraint demands the late-time potential density of the scalar field to be 
extremely small. This huge difference of energy scales (of order 
${\cal O}(10^{100})$!) is the basis for the $\eta$-problem of quintessential 
inflation. 

\subsection{\boldmath The $\eta$-problem of quintessential inflation}

In conventional cosmology, in order for inflation to last enough 
e-foldings to account for the horizon problem without an initially 
super-Planckian $V^{1/4}$, it is necessary for the potential to be rather flat
during inflation. As a result, in order to prepare for the abysmal ``dive'' 
after the end of inflation (so as to cover the huge difference of energy 
scales) the curvature $V''$ of the potential near the end of inflation is
substantial. As a result the spectral index $n_s$ of the inflation--generated
density perturbation spectrum is too large compared with the observational 
requirement:

\begin{equation}
|n_s-1|\leq 0.1
\label{nscons}
\end{equation}

This can be understood from the fact that $n_s$ is given by \cite{book}

\begin{equation}
n_s=1+2(\eta-3\epsilon)
\label{ns}
\end{equation}
where $\eta$ and $\epsilon$ are the so--called slow roll parameters of 
inflation, which, in conventional cosmology, are defined as

\begin{eqnarray}
\epsilon\equiv-\frac{\dot{H}}{H^2}\simeq
\frac{m_P^2}{2}\left(\frac{V'}{V}\right)^2
 & \qquad {\rm and} \qquad &
\eta\equiv m_P^2\frac{V''}{V}
\label{sr}
\end{eqnarray}

\begin{figure}
\begin{center}
\leavevmode
\hbox{
\epsfxsize=3.9in
\epsffile{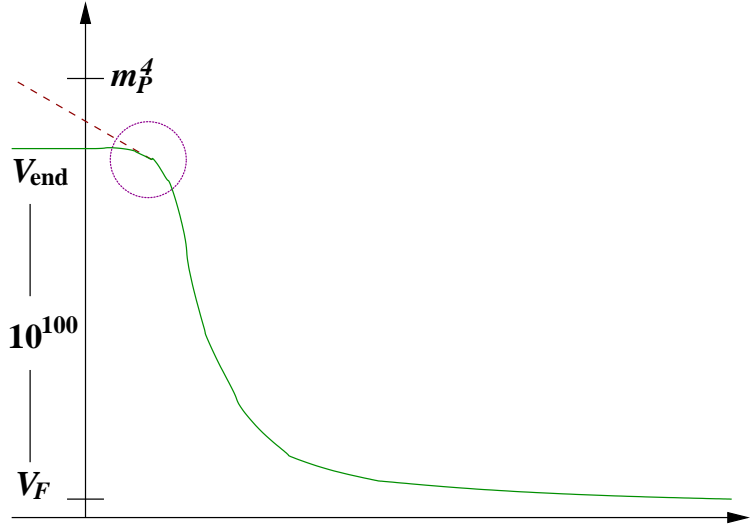}}
\end{center}
\caption{\footnotesize
Graphic illustration of the source of the $\eta$-problem of quintessential
inflation. In order to bridge the huge gap between the energy scales 
corresponding to the inflationary plateau and the quintessential tail the
curvature of the potential near the end of inflation is too large to allow
for an almost scale invariant spectrum of curvature perturbations. An attempt
to steepen the inflationary plateau, however, would reduce the total number of 
inflationary e-foldings below the necessary amount required for the solution
of the horizon problem, unless the inflationary scale was allowed to 
superseded the Planck scale.
}
\label{etafig}
\end{figure}

Therefore, a strongly curved potential results in unacceptably large $|\eta|$,
which, in turn, because of (\ref{ns}), causes deviations from scale invariance 
that are incompatible with observations. An illustration of the $\eta$-problem 
can be seen in Figure~\ref{etafig}.

The hope had been that brane-cosmology, since it allows overdamped steep
inflation, would be able to avoid a strongly curved inflationary 
potential without introducing super-Planckian densities. This is because, as 
evident by (\ref{Hfric}), for energies above the brane tension $\lambda$, the 
Hubble parameter is larger than in the usual FRW case. This introduces extra 
friction in the roll--down of the scalar field, as determined by (\ref{field}). 
Consequently, the roll becomes much slower and, even with sufficiently large
number of inflationary e-foldings to solve the horizon problem, the field 
rolls so little that super-Planckian densities can be avoided. Moreover, 
slow-roll can be achieved even when dispensing with the inflationary plateau,
leading to the so--called steep inflation \cite{steep}, which again assists 
in reducing the curvature of the potential during inflation. 

\subsection{Coincidence and BBN}\label{coincbbn}

However, as we show below, the above beneficial effects of brane-cosmology are
counteracted by the consequences of extra friction in the period of kination.
Indeed, overdamped kination is reduced in duration. As a result, the field
is not able to roll as much down its quintessential tail as it would
in conventional cosmology, which intensifies the already stringent constraints
of coincidence and BBN. To demonstrate this in a quantified way, we study below
these constrains considering an exponential quintessential tail of the form:

\begin{equation}
V(\phi)=V_{\rm end}\exp(-b\Delta\phi/m_P)
\end{equation}
where \mbox{$\Delta\phi\equiv\phi-\phi_{\rm end}$}. Using (\ref{bfF}) the
coincidence requirement (\ref{coinc0}) results in the constraint:

\begin{equation}
\begin{array}{l}
\left(1\!+\!\sqrt{\frac{3}{2}}\,b\right)
\ln\!\Big(\mbox{\large $\frac{m_P^4}{V_{\rm end}}$}\Big)
\mbox{=}\ln\!\Big(\mbox{\large $\frac{m_P^4}{\rho_0}$}\Big)
\mbox{-} \ln\Omega_\phi \,\mbox{-}
\sqrt{\frac{3}{2}}\,b\!\left[\frac{8}{3}\!+\!
2\ln\!\left(\!\frac{48\pi}{\alpha^2}\!\sqrt{\frac{30}{g_{\rm reh}}}\right)
\mbox{-} \frac{4}{3}\sqrt{\mbox{\large $\frac{\lambda}{2V_{\rm end}}$}}\, 
\mbox{-}
\frac{7}{3}\ln\!\left(\mbox{\large $\frac{2V_{\rm end}}{\lambda}$}\right)
\right]
\end{array}
\label{coinc}
\end{equation}

Similarly, the requirement \mbox{$T_*>T_{\sc bbn}$}, in view of (\ref{bT*}),
becomes

\begin{equation}
\ln\!\left(\frac{m_P^4}{V_{\rm end}}\right)\leq
\ln\!\left[\frac{\alpha^3}{(24\pi)^2}\sqrt{\frac{g_{\rm reh}}{10}}
\left(\frac{g_{\rm reh}}{g_*}\right)^{1/4}\right]
+\frac{3}{2}\ln\!\left(\frac{2V_{\rm end}}{\lambda}\right)
+\ln\!\left(\frac{m_P}{T_{\sc bbn}}\right)
\label{bbn}
\end{equation}

Combining the above one finds the bound: \mbox{$b\geq b_{\rm min}$}, where

\begin{equation}
b_{\rm min}\equiv\sqrt{\frac{2}{3}}\;\;
\frac{
\ln\!\Big(\mbox{\large $\frac{m_P^4}{\rho_0}$}\Big)-\ln\Omega_\phi-
\ln\!\left[\mbox{\large $\frac{\alpha^3}{(24\pi)^2}$}
\sqrt{\mbox{\large $\frac{g_{\rm reh}}{10}$}}
\left(\mbox{\large $\frac{g_{\rm reh}}{g_*}$}\right)^{1/4}\right]
-\ln\!\left(\mbox{\large $\frac{m_P}{T_{\sc bbn}}$}\right)-
\mbox{\large $\frac{3}{2}$}
\ln\!\left(\mbox{\large $\frac{2V_{\rm end}}{\lambda}$}\right)
}{
\ln\!\left[\mbox{\large $\frac{12}{\alpha}$}
\sqrt{\mbox{\large $\frac{10}{g_{\rm reh}}$}}
\left(\mbox{\large $\frac{g_{\rm reh}}{g_*}$}\right)^{1/4}\right]
+\mbox{\large $\frac{8}{3}$}+
\ln\!\left(\mbox{\large $\frac{m_P}{T_{\sc bbn}}$}\right)-
\mbox{\large $\frac{4}{3}$}\sqrt{\mbox{\large $\frac{\lambda}{2V_{\rm end}}$}}
-\mbox{\large $\frac{5}{6}$}
\ln\!\left(\mbox{\large $\frac{2V_{\rm end}}{\lambda}$}\right)
}
\end{equation}

The above expression looks rather complicated but, in fact, it becomes quite
simple once the numbers are introduced. We are going to use
\mbox{$g_*=10.75$} and \mbox{$g_{\rm reh}=106.75c$}, where \mbox{$c=1$} for
the SM and \mbox{$c\gsim 2$} for its supersymmetric extensions. Also,
in order to compensate for the overestimate of $T_*$ in (\ref{bT*}), we
will use \mbox{$T_{\sc bbn}\simeq 10$ MeV}. Finally, let us 
define: \mbox{$Y\equiv\ln(2V_{\rm end}/\lambda)$}. Then we find

\begin{equation}
b_{\rm min}=\sqrt{\frac{2}{3}}\;
\frac{236.66-\frac{3}{4}\ln c-3\ln\alpha-\frac{3}{2}Y}{
51.48-\frac{1}{4}\ln c-\ln\alpha-\frac{4}{3}e^{-Y/2}-\frac{5}{6}Y}
\label{bbmin}
\end{equation}

Thus we see that $b_{\rm min}$ grows with $Y$, which means that the more
prominent the brane effect becomes the more the parameter space for $b$
shrinks. Indeed, remember that, from (\ref{brange}), the maximum acceptable
value of $b$, in order for brief acceleration to occur, is 
\mbox{$b_{\rm max}=2\sqrt{6}$}. 

Thus, as far as kination and BBN are concerned, the case of conventional 
cosmology is preferable. We can recover conventional cosmology if we set
\mbox{$Y=0$} and \mbox{$\alpha\rightarrow 2\alpha$}. This gives

\begin{equation}
b_{\rm min}=\sqrt{\frac{2}{3}}\;
\frac{234.58-\frac{3}{4}\ln c-3\ln\alpha}{
49.46-\frac{1}{4}\ln c-\ln\alpha}
\label{bmin}
\end{equation}

The lowest value for the above corresponds to \mbox{$c=1$} and
\mbox{$\alpha\simeq 0.1$}, for which we find \mbox{$b_{\rm min}=3.81$}.
Note that, in both conventional and brane cosmology, 
\mbox{$b_{\rm min}>b_{\rm min}(\alpha\rightarrow 0)=\sqrt{6}$}. According to 
\cite{exp}, when \mbox{$b\geq\sqrt{6}$} the attractor (\ref{fattr})
is unreachable. Instead, after unfreezing the field engages again in free-fall 
evolution, where \mbox{$\rho_\phi\simeq\rho_{\rm kin}\propto a^{-6}$}, until 
it refreezes at another value \mbox{$\phi_{F'}=\phi_F+\sqrt{2V_F}\,t_F$}, 
where $t_F$ is the time of unfreezing. This is true regardless of $w_B$ as can 
be shown easily through the use of (\ref{fieldkin}). Using (\ref{evol}) we find

\begin{equation}
\phi_{F'}=\phi_F+\frac{4}{\sqrt{6}(1+w_B)}\,m_P
\end{equation}

\begin{figure}
\begin{center}
\leavevmode
\hbox{
\epsfxsize=5in
\epsffile{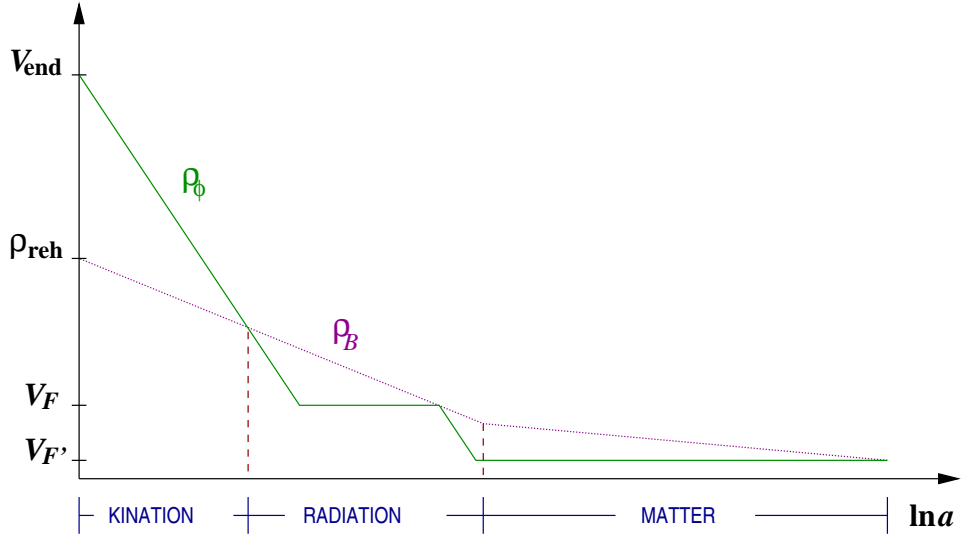}}
\end{center}
\caption{\footnotesize
The evolution of $\rho_\phi$ (solid line) and $\rho_B$ (dotted line) in the
multiple--unfreezings scenario. For \mbox{$b\geq\sqrt{6}$}, after unfreezing 
from $V_F$ the field engages again into free fall until its kinetic energy is 
depleted once more, when it freezes again at a new position with potential 
density \mbox{$V_{F'}\equiv V(\phi_{F'})\ll V_F$}. The process may be 
repeated many times. Each unfreezing stage causes a brief period of 
accelerated expansion for the Universe and also sends $\phi$ further down
its potential so that coincidence becomes easier to achieve. 
}
\label{kin2fig}
\end{figure}

The process can be repeated again 
and again, leading to many `glitches' of brief accelerated expansion. This 
effect may enlarge the parameter space since it is expected to relax the 
coincidence constraint because the final value of $V$ today can be lower than 
$V_F$. This possibility certainly deserves further investigation, which, 
however, we will not pursue here. An illustration of this process is shown
in Figure~\ref{kin2fig}.

\subsection{Gravitational waves}

Another important constraint related to the kination period has to do
with the spectrum of the Gravitational Waves (GW) generated during inflation.
Because of the stiffness of the equation of state of the Universe, the 
GW~spectrum forms a spike at high frequencies, instead of being flat as is the
case for the radiation era \cite{giova}. 
Indeed, it has been shown that the GW~spectrum is
of the form \cite{giova}\cite{sss}:

\begin{equation}
\Omega_{\sc gw}(k)=\left\{
\begin{array}{ll}
\varepsilon\Omega_\gamma(k_0)h_{\sc gw}^2
\left(\mbox{\large $\frac{k}{k_*}$}\right)
[\ln(k/k_{\rm end})]^2 & k_*<k\leq k_{\rm end}\\
 & \\
\frac{\pi}{4}\varepsilon\Omega_\gamma(k_0)h_{\sc gw}^2
[\ln(k_*/k_{\rm end})]^2 & k_{\rm eq}<k\leq k_*\\
 & \\
\frac{\pi}{16}\varepsilon
\Omega_\gamma(k_0)h_{\sc gw}^2
\left(\mbox{\large $\frac{k_{\rm eq}}{k}$}\right)^2
[\ln(k_*/k_{\rm end})]^2 & k_0<k\leq k_{\rm eq}
\end{array}\right.
\label{GWspec}
\end{equation}
where $\Omega_{\sc gw}(k)$ is the density fraction of the gravitational waves
with {\em physical} momentum $k$, 
\mbox{$\Omega_\gamma(k_0)=2.6\times 10^{-5}h^{-2}$} is the density fraction
of radiation at present on horizon scales (\mbox{$h=0.65$} is the Hubble 
constant $H_0$ in units of 100 km/sec/Mpc) and the subscripts `$*$' and `eq' 
denote the end of kination (onset of radiation era) and the end of radiation 
era (onset of matter era) respectively. Moreover, 
\mbox{$\varepsilon=\alpha_{\sc gw}/2\pi\sim 10^{-2}$} with 
\mbox{$\alpha_{\sc gw}\sim 0.1$} being the GW~generation efficiency during 
inflation and \cite{sss}\cite{lang}

\begin{equation}
h_{\sc gw}\equiv\frac{1}{\pi}\left(\frac{H_{\rm end}}{m_P}\right)
{\cal F}(\rho/\lambda)
\end{equation}
where \cite{lang}
\begin{equation}
{\cal F}(x)\equiv\left[\sqrt{1+x^2}-
x^2\ln\left(\frac{1}{x}+\sqrt{1+\frac{1}{x^2}}\right)\right]^{-1/2}
\!\!\Rightarrow\; {\cal F}(\rho/\lambda)\simeq\left\{
\begin{array}{ll}
1 & \rho\ll\lambda\\
\sqrt{\frac{3}{2}(\rho/\lambda)} & \rho\gg\lambda
\end{array}\right.
\end{equation}

The danger is that the generated GWs may destabilize BBN. The relevant 
constraint on $\Omega_{\sc gw}(k)$ reeds:

\begin{equation}
I\equiv h^2\int_{k_{\sc bbn}}^{k_{\rm end}}\Omega_{\sc gw}(k)\;
d\ln k\leq 2\times 10^{-6}
\label{bbnGW}
\end{equation}
where $k_{\sc bbn}$ is the physical momentum that corresponds to the horizon
at BBN. From (\ref{GWspec}) it is easy to find

\begin{equation}
I=h^2\varepsilon\Omega_\gamma(k_0)h_{\sc gw}^2
\left\{2\left(\frac{k_{\rm end}}{k_*}\right)-
\left[\ln\left(\frac{k_{\rm end}}{k_*}\right)+1\right]^2+
\frac{\pi}{4}
\left[\ln\left(\frac{k_{\rm end}}{k_*}\right)\right]^2
\ln\left(\frac{k_*}{k_{\sc bbn}}\right)\right\}
\end{equation}

Since \mbox{$k_{\rm end}\gg k_*>k_{\sc bbn}$} 
the expression in brackets above is dominated by the first term. 
We also have

\begin{equation}
\frac{k_{\rm end}}{k_*}=\frac{H_{\rm end}}{H_*}
\left(\frac{a_{\rm end}}{a_*}\right)=
\left(\frac{H_{\rm end}}{H_*}\right)^{2/3}
\left(\frac{H_{\rm end}}{H_\lambda}\right)^{1/6}
\end{equation}
where \mbox{$H_\lambda\equiv H(t_\lambda)$} and the last factor reduces to
unity when considering conventional kination. Putting all these together we
find

\begin{equation}
I=2h^2\varepsilon\Omega_\gamma(k_0)h_{\sc gw}^2
\left(\frac{H_{\rm end}}{H_*}\right)^{2/3}
\left(\frac{H_{\rm end}}{H_\lambda}\right)^{1/6}
\end{equation}

Inserting the above into (\ref{bbnGW}) and after some algebra we end up with 
the constraint:

\begin{equation}
\alpha^4\geq\frac{15}{g_{\rm reh}}
\frac{\lambda{\cal F}^2(V_{\rm end}/\lambda)}{(m_PH_{\rm end})^2}
\left(\frac{2V_{\rm end}}{\lambda}\right)^{3/2}
\label{GW0}
\end{equation}

In the case of brane cosmology we have 
\mbox{${\cal F}^2\simeq\frac{3}{2}(V_{\rm end}/\lambda)$} so that the bound 
becomes

\begin{equation}
\alpha\gsim\left(\frac{270}{g_{\rm reh}}\right)^{1/4}
\left(\frac{2V_{\rm end}}{\lambda}\right)^{1/8}
\label{bGW}
\end{equation}
whereas for conventional cosmology \mbox{${\cal F}^2\simeq 1$} and the bound is

\begin{equation}
\alpha\gsim\left(\frac{90}{g_{\rm reh}}\right)^{1/4}
\label{GW}
\end{equation}

Thus, we see that the brane-effect also sets a somewhat tighter lower bound on 
the reheating efficiency due to excessive GW~generation. In both cases 
\mbox{$\alpha\gsim 1$} and, therefore, purely gravitational reheating is only
marginally compatible with the GW~constraint.

Here, we should mention another, potentially more dangerous relic, introduced
by gravitational reheating, namely gravitinos. Gravitino overproduction is also
possible to endanger BBN. In fact they are rather stringently constrained as
\cite{gravitini}

\begin{equation}
\frac{n_g}{s}\leq 10^{-14}
\end{equation}
where $n_g$ is the number density of the gravitinos which is kept in constant
ratio with the entropy $s$ of the Universe. The above ratio is easy to compute
\cite{inf}:

\begin{equation}
\frac{n_g}{s}=\frac{135\zeta(3)}{2\pi^4g_{\rm reh}}
\left(\frac{\alpha_g}{\alpha}\right)^3
\end{equation}
where \mbox{$\zeta(3)=1.20206$} and $\alpha_g$ is the production efficiency of 
gravitinos. The above provide the following lower bound on the reheating
efficiency:

\begin{equation}
\alpha\geq 9\times 10^3c^{-1/3}\alpha_g
\label{ag}
\end{equation}
According to \cite{gravitini} gravitino production can be as efficient as 
the gravitational production of any other particle, i.e. 
\mbox{$\alpha_g\sim 0.1$}, even though {\em the gravitinos are not generated
during inflation but only afterwards} (that is at the end of inflation). 
Indeed, the gravitino overproduction danger concerns the spin-$\frac{1}{2}$
gravitinos and not the usual spin-$\frac{3}{2}$ ones. The spin-$\frac{1}{2}$
gravitinos (longitudinal modes) are massive because they absorb the goldstino
mode and this is why they cannot be generated {\em during} inflation. Still,
to date there is no thorough calculation of $\alpha_g$ in a stiff equation 
of state and also in the case of brane--cosmology so, the gravitino bound 
(\ref{ag}) may not be as reliable as the bounds due to GW~generation.

In a similar way as described above, the stiff equation of state during
kination may lead to efficient production of supersymmetric dark matter, e.g. 
neutralinos \cite{salati}. Moreover, the fluctuations of the inflaton field 
itself can be considered as dark matter \cite{DM}. Finally, if the rolling 
scalar field is even weakly coupled to SM fields it may lead to 
substantial leptogenesis or baryogenesis even though the Universe is in 
thermal equilibrium, which may explain the observed baryon asymmetry 
\cite{baryon}. It has been shown that the backreaction of the latter effect 
does not affect the dynamics of $\phi$ and (\ref{field}) is still valid.

\section{The curvaton hypothesis}

As we have shown in the previous section, even though brane cosmology may 
help with the $\eta$-problem by allowing overdamped steep inflation, it is
this very effect of overdamping that turns negative during kination by making 
it harder for the field to roll down enough so as to achieve successful 
coincidence. Is, then, all lost for quintessential inflation?

Fortunately it is not.
An alternative way to ameliorate the $\eta$-problem is through the so--called
curvaton hypothesis \cite{curv}. According to this hypothesis the curvature 
perturbation spectrum, which seeds the formation of Large Scale Structure and 
the observed anisotropy of the Cosmic Microwave Background Radiation (CMBR), 
is due to the amplification of the quantum fluctuations of a scalar field 
{\em other than the inflaton} during inflation. This field $\sigma$, called
curvaton, has to satisfy certain requirements to fulfill its role in generating
the correct curvature perturbation spectrum. In order for its quantum 
fluctuations to get amplified during inflation the curvaton $\sigma$, much 
like the inflaton in conventional inflation, has to be an effectively massless 
scalar field, with mass \mbox{$m_\sigma<\frac{3}{2}\,H_{\rm inf}$}, where 
$H_{\rm inf}$ is the Hubble parameter during inflation. Also, in order for the 
generated perturbations to be Gaussian, in accordance to observations, the 
curvaton should be significantly displaced from its vacuum expectation value 
(VEV) during inflation, i.e. 
\mbox{$|\sigma-\langle\sigma\rangle|\gg H_{\rm inf}$}. 
However, the curvaton's contribution to the potential density during inflation 
is negligible and this is why inflationary dynamics is still governed by the 
inflaton field. One final requirement for a successful curvaton field is that 
its couplings to the reheated thermal bath are small enough to prevent its 
thermalization after the end of inflation (which would, otherwise, wipe out 
its superhorizon perturbation spectrum). 

The curvaton, being subdominant and effectively massless during inflation 
remains overdamped and, more or less, frozen. After the end of inflation
$\sigma$ remains frozen until \mbox{$H(t)\sim m_\sigma$}, when the field 
unfreezes and begins oscillating around its VEV. Doing so its average energy 
density scales as pressureless matter, i.e. 
\mbox{$\rho_\sigma\propto a^{-3}$}. This means that, if the unfreezing of 
the curvaton occurs early enough (i.e. before the matter era) the latter comes
to dominate the Universe, causing a brief period of matter domination, until 
it decays into a new thermal bath comprised by the curvaton's decay 
products. This is expected to somewhat relax the GW and gravitino constraints 
because the additional entropy production by the decay of the curvaton will 
dilute the 
GW/gravitino contribution to the overall density.\footnote{It is also possible 
for the curvaton to decay just before it dominates the Universe, which allows 
a certain isocurvature component in the density perturbations.} The curvature 
perturbation spectrum of $\sigma$ is imposed onto the Universe, when the 
latter becomes curvaton dominated (or nearly dominated). 

There are two important differences between the curvaton hypothesis and 
conventional inflation. Firstly, because the curvature perturbation spectrum
is due to the curvaton the spectral index is not given by (\ref{ns}) but by
\cite{curv}

\begin{equation}
n_s=1+2(\eta_{\sigma\sigma}-\epsilon\,)
\end{equation}
where 

\begin{equation}
\eta_{\sigma\sigma}\equiv \frac{m_P^2}{V}\frac{\partial^2V}{\partial\sigma^2}
\end{equation}

Now, since the $\sigma$-dependent part of $V$ is not related to inflation
$\eta_{\sigma\sigma}$ can be extremely small. This means that the spectral 
index constraint (\ref{nscons}) becomes

\begin{equation}
\epsilon<0.05
\label{epscons}
\end{equation}
which is possible to satisfy even for large $\eta$ and much easier too.
Thus, {\em the $\eta$-problem for quintessential inflationary model building
is ameliorated through the curvaton hypothesis because one can keep an almost 
scale invariant spectrum of curvature perturbations even with a substantially 
curved scalar potential}.

The second effect of the curvaton hypothesis on inflationary model building
is the fact that the {\sc cobe} observations impose {\em only an upper bound} 
on the amplitude of the inflaton generated curvature perturbations. If we want 
to allow for a large $\eta$ then this bound should be

\begin{equation}
\frac{1}{2\pi}\left.\frac{\delta\phi}{\phi}\right|_{\rm exit}\leq 0.1 
\left(\frac{\Delta T}{T}\right)_{\sc cobe}\simeq 5\times 10^{-6}
\end{equation}
which, for slow roll inflation, can be recast as

\begin{equation}
\frac{1}{\sqrt{3}\pi}\frac{V^{3/2}}{m_P^3|V'|}\leq 10^{-5}
\label{cobe}
\end{equation}

There are numerous candidates for successful curvatons, especially in 
supersymmetric theories, where scalar fields are abundant. Of particular 
interest are pseudo-Goldstone bosons or axion-like string moduli, because
their mass is protected by symmetries and can be rather small during 
inflation. In \cite{mine} the liberation effect of the curvaton hypothesis
on inflationary model building has been shown by demonstrating how it can
rescue a number of, otherwise unviable inflationary models, which are well 
motivated by particle physics.

In the following sections we will apply the curvaton hypothesis on 
quintessential inflation model building both in conventional and brane 
cosmology, demonstrating thereby the fact that the $\eta$-problem is, indeed, 
substantially ameliorated.

\section{The case of Standard Cosmology}

Let us first consider the case of conventional cosmology, where kination is
not inhibited by overdamping effects. We focus on modular inflation which 
has the merit that the scalar field is a modulus, which corresponds to a flat 
direction protected from excessive supergravity corrections and may refrain 
from steepness even when the field travels distances as large as $m_P$ in 
field space, a problem which, in most models of quintessence, is unresolved 
\cite{KL}.

\subsection{Modular inflation}\label{modular}

Moduli fields correspond to flat directions in field space that are protected 
by symmetries against supergravity corrections. However, the values of 
string-inspired moduli are typically related to observables, such as the gauge 
coupling in the case of the dilaton, and need to become stabilized. This is 
usually supposed to occur at inner-space distances of order $m_P$, where 
non-perturbative K\"{a}hler corrections may generate a minimum for the field. 
Thus, the expected VEV for a modulus is \mbox{$\langle\phi\rangle\sim m_P$}. 
Therefore, the scalar potential for a modulus near its origin would be

\begin{equation}
V(\phi)=V_0-\frac{1}{2}m^2\phi^2+\cdots
\label{Vmodul}
\end{equation}
where, $V$ is expected to depart from $V_0$ when $\delta\phi\sim m_P$, so that

\begin{equation}
V_0\sim (m_Pm)^2
\label{V0}
\end{equation}

The inflationary scale is usually taken to be the so--called intermediate scale
\mbox{$V_0^{1/4}\simeq 5\times 10^{10}$} corresponding to gravity mediated 
supersymmetry breaking. Then, from the above \mbox{$m\sim 1$ TeV}. As a result
we find

\begin{equation}
|\eta|=\frac{(m_Pm)^2}{V_0}\sim 1
\label{etamodul}
\end{equation}
which means that such a modulus field cannot be the inflaton of conventional
inflation because it would be impossible to attain a scale invariant spectrum
of curvature perturbations. Moreover, the inflationary energy scale is too 
low to generate the necessary amplitude for the curvature perturbations.

In contrast, as shown in \cite{mine}, modular inflation works fine in the 
context of the curvaton hypothesis. Indeed, from (\ref{Vmodul}), it is easy to 
see that

\begin{equation}
\epsilon=\frac{\eta^2}{2}\left(\frac{\phi}{m_P}\right)^2
\end{equation}
which can become very small near the origin and easily satisfy the constraint
(\ref{epscons}). The question is, of course, why should $\phi$, stand at the 
origin in the first place. This is natural to occur if the origin is point of 
enhanced symmetry \cite{enhanced}, where the modulus field has strong 
couplings with the fields of some thermal bath preexisting inflation. 
Such strong couplings
introduce temperature corrections to (\ref{Vmodul}) which drive $\phi$ 
to zero. The inflationary expansion, then, begins with a period of thermal
inflation, which inflates away the primordial thermal bath and renders the 
origin a local maximum. Afterwards, quantum fluctuations send the field 
rolling down and away from the origin, in a period of fast-roll inflation. 
This model, called thermal modular inflation, is discussed in 
\cite{mine}.\footnote{However, we do not need to presuppose so much. In fact 
one can use anthropic-style arguments and consider the fact that only patches 
of the Universe where $\phi$ is near the origin will inflate (the nearer the 
more inflation) and, therefore, the likelihood to be living in one of them is 
greatly enlarged.}

It is possible to formulate a model of quintessential inflation based on 
modular inflation if one considers that the supergravity corrections 
introduced into the potential at \mbox{$\phi\sim m_P$} may not generate
a minimum for the potential but just give rise to a slope, with the minimum
displaced at infinity. After all, for the moduli one only expects
that \mbox{$\delta V(\delta\phi\sim m_P)\sim V$}. Thus, for example, the 
potential may look like this

\begin{equation}
V(\phi)\simeq\left\{
\begin{array}{lc}
V_0-\frac{1}{2}m^2\phi^2 & 0<\phi\ll m_P\\
 & \\
V_0\exp(-b\phi/m_P) & \phi\gg m_P
\end{array}\right.
\label{modqinf0}
\end{equation}

This form is rather plausible for moduli potentials. Indeed, 
the $F$-term scalar potential in supergravity is

\begin{equation}
V\simeq e^{K/m_P^2}|W|^2\left[\sum_{nm}
\left(\frac{K_n}{m_P^2}+\frac{W_n}{W}\right)K^{nm}
\left(\frac{K_m}{m_P^2}+\frac{\bar{W}_m}{\bar{W}}\right)
-3\,m_P^{-2}\right]
\end{equation}
where $W$ is the superpotential, $K$ is the K\"{a}hler potential, 
\mbox{$K^{nm}=(K_{nm})^{-1}$}, the overbar denotes charge conjugation
and the subindices represent derivatives with respect to the different
fields of the theory. In many string models the dynamics of the above is 
dominated by the $e^{K/m_P^2}$ factor (see for example 
\cite{decarlos1}). Now, the K\"{a}hler potential, at tree level, is 
logarithmic with respect to the moduli $\Phi_i$ such that 
\mbox{$K\propto -m_P^2\sum_i\ln[(\Phi_i+\bar{\Phi}_i)/m_P]$}, 
which means that 
\mbox{$e^{K/m_P^2}\propto 1/\prod_i[(\Phi_i+\bar{\Phi}_i)/m_P]$}. 
Note that the $\Phi_i$ moduli do not have canonical kinetic terms. Instead the 
kinetic part of the relevant Lagrangian density is given by

\begin{equation}
{\cal L}_{\rm kin}=%\frac{1}{2}\,
K_{ij}\;\partial_\mu\Phi_i\partial^\mu\bar{\Phi}_j
\end{equation}
which means that we can define the canonically normalized moduli as 
\mbox{$\phi_i\propto\ln[(\Phi_i+\bar{\Phi}_i)/m_P]$}, in terms of which the 
scalar potential becomes an exponential, i.e. 
\mbox{$V\propto\exp(-\sum_ib_i\phi_i/m_P)$}.
the values of the positive $b_i$ coefficients in the exponents depend on the 
particular string model considered but, in general they are of order unity
(for example in \cite{decarlos1} \mbox{$b=2\sqrt{2}$} whereas in 
\cite{mazum} \mbox{$b=4\sqrt{\pi}$}). Obviously, the potential is eventually
dominated by the term with the smallest $b_i$. 

The potential can easily form a maximum at the origin if there exists
a discrete symmetry of the form \mbox{$\phi_i\rightarrow-\phi_i$} 
(which corresponds to the well known $T$-duality: 
\mbox{$e^{\phi_i}=1/e^{\phi_i}$}). In this case the couplings of the 
moduli with matter at the origin are maximized \cite{duality}, exactly
as required by thermal modular inflation. In contrast, away from the origin,
these couplings are strongly suppressed leading to an effectively sterile 
inflaton, as required by quintessential inflation. 

It is important to note that in the case described above the modulus is not 
stabilized by reaching its VEV, but it does so dynamically, when 
reaching \mbox{$\phi=\phi_F$} where it freezes. 
Of course $\phi_F$ has to be at the correct value for phenomenology to
work. This is especially true for the dilaton, which determines the 
gauge coupling. Thus, it would be safer to consider the so-called geometrical 
moduli ($T$-moduli) associated with the volume of the extra dimensions. The 
dependence of the SM--phenomenology on these is not manifest at tree-level but 
arises only at one-loop and beyond. 

The above are based on the implicit assumption that the superpotential has 
only a weak dependence on the moduli and, therefore, $V$ is mostly determined 
by the $e^{K/m_P^2}$ factor. However, it should be pointed out here that, 
according to the usual interpretation of (heterotic) string phenomenology,
the superpotential receives non-perturbative contributions from hidden sector 
gaugino condensates, which are of the form 
\mbox{$W\propto\exp(-\sum_i\beta_i\Phi_i/m_P)$}. Consequently, a $T$-modulus 
would have a double exponential potential. As discussed in \cite{jose}, such a 
potential, being steeper than the pure exponential, has a disastrous 
attractor solution. Indeed, not only does this attractor diminish 
$\rho_\phi$ much faster than $\rho_B$ but it is also attained very soon after 
the end of inflation and, therefore, renders late-time $\phi$-domination 
impossible. However, not all the possibilities for the moduli have been 
explored and there are more types of string theory than the usual heterotic 
string. So, we believe that it is quite possible that a canonically normalized
modulus may have a scalar potential with the desired pure exponential tail.

Below we will examine the bahavour of a toy model that bares the 
characteristics outlined above and investigate whether it is indeed 
possible to be a successful quintessential inflationary model. We name this
proposal modular quintessential inflation.

\subsection{Modular quintessential inflation}

\subsubsection{The toy model}

Consider the potential:

\begin{equation}
V(\phi)=\frac{M^4}{[\cosh(\phi/m_0)]^q}
\label{cosh}
\end{equation}
where $q$ is a positive integer and $M,m_0$ are mass scales. The above becomes

\begin{equation}
V(\phi)\simeq\left\{
\begin{array}{lc}
M^4-\frac{1}{2}q(M^2/m_0)^2\phi^2 & \qquad 0<\phi\ll m_P\\
 & \\
2^qM^4\exp(-q\phi/m_0) & \qquad \phi\gg m_P
\end{array}\right.
\label{modqinf}
\end{equation}
which can be identified with (\ref{modqinf0}) if we define:

\begin{eqnarray}
m^2\equiv q\,\frac{M^4}{m_0^2} &
\qquad V_0\equiv M^4 \qquad &
b\equiv q\,\frac{m_P}{m_0}
\label{scales}
\end{eqnarray}

The slow roll parameters for the above model are

\begin{eqnarray}
\epsilon=\frac{1}{2}\,b^2[\tanh(\phi/m_0)]^2
 & \qquad &
\eta=2\epsilon-\frac{b^2}{q}[\cosh(\phi/m_0)]^{-2}
\label{etaeps}
\end{eqnarray}
 In order to have enough e-foldings of inflation we need
\mbox{$\eta(\phi\rightarrow 0)\geq -1$}, which demands \mbox{$q\geq b^2$}. Then
it can be shown that inflation ends at

\begin{equation}
\phi_{\rm end}\simeq\sqrt{\frac{2}{|\eta|}}\,m_P\simeq\frac{\sqrt{2q}}{b}\,m_P
\label{fend}
\end{equation}

The number of fast-roll e-foldings before the end of inflation is related to
the value of the scalar field at that time $\phi_N$ by \cite{fastroll}

\begin{equation}
N\simeq\frac{1}{F}
\ln\left(\frac{\phi_{\rm end}}{\phi_N}\right)
\Rightarrow \phi_N=\phi_{\rm end}\exp(-FN)
\label{N}
\end{equation}
where 

\begin{equation}
F\equiv\frac{3}{2}\left(\sqrt{1+\frac{4}{3}|\eta|}-1\right)
\end{equation}
which, for slow roll inflation, becomes \mbox{$F(|\eta|\ll 1)\approx|\eta|$}.

\subsubsection{Enforcing the constraints}

Let us employ now the {\sc cobe} bound (\ref{cobe}). We find that the 
bound translates into a lower bound on $q$ such that 
\mbox{$q\geq q_{\rm min}$}, where

\begin{equation}
q_{\rm min}\equiv
\frac{b^2N_{\rm dec}}{2\mu-5\ln10+\frac{1}{2}\ln(6\pi^2/q_{\rm min})+\ln b}
\label{qmin}
\end{equation}

In the above we have defined \mbox{$\mu\equiv\ln(m_P/M)$} and also
$N_{\rm dec}$ is the number of inflationary e-foldings that
remain when the scale, which reenters the horizon at decoupling (corresponding
to the time of emission of the CMBR), exits the horizon during inflation.
This scale is related to the reheating efficiency by \cite{jose}

\begin{equation}
N_{\rm dec}=\ln(T_{\sc cmb}t_0)+\ln(H_{\rm end}/T_{\rm reh})=
66.94-\ln\alpha
\label{Ndec}
\end{equation}
where $T_{\sc cmb}$ is the temperature of the CMBR at the present time $t_0$.

Let us now enforce the coincidence constraint (\ref{coinc}) in the case
of conventional cosmology (\mbox{$\lambda=2V_{\rm end}$} and 
\mbox{$\alpha\rightarrow 2\alpha$}). With a little algebra we find

\begin{equation}
b=\sqrt{\frac{2}{3}}\frac{69.18-\mu}{1.83-\frac{1}{4}\ln c-\ln\alpha+\mu}
\label{coinc1}
\end{equation}
which diminishes with $\mu$ and, therefore, we can define
\mbox{$\mu_{\rm min}\equiv\mu(b_{\rm max})$}, where, according to 
(\ref{brange}), \mbox{$b_{\rm max}=2\sqrt{6}$}. Thus, we obtain

\begin{equation}
\mu_{\rm min}=8.31+\frac{3}{14}\ln c+\frac{6}{7}\ln\alpha
\label{mmin}
\end{equation}

Finally, let us use the BBN constraint (\ref{bbn}) to obtain the upper bound on
$\mu$. Similarly as above we find

\begin{equation}
\mu_{\rm max}=10.53+\frac{3}{16}\ln c+\frac{3}{4}\ln\alpha
\label{mmax}
\end{equation}

Both $\mu_{\rm min}$ and $\mu_{\rm max}$ increase with $\alpha$, but with
different rates so that the $\mu$-range decreases. Thus there is an upper bound
on $\alpha$ where \mbox{$\mu_{\rm min}=\mu_{\rm max}$}. It is easy to see that 

\begin{equation}
\ln\alpha_{\rm max}=20.72-\frac{1}{4}\ln c\Rightarrow
\alpha_{\rm max}\approx 10^9c^{-1/4}
\label{amax}
\end{equation}

The lower bound on the reheating efficiency is set by the GW~constraint 
(\ref{GW}). Therefore, the $\alpha$-range is

\begin{equation}
1\leq\alpha\leq 10^9
\label{arange}
\end{equation}

It can be checked that even $\alpha_{\rm max}$ is much smaller that the
reheating efficiency $\alpha_{\rm pr}$, which corresponds to prompt reheating: 
\mbox{$\rho_{\rm reh}(\alpha_{\rm pr})=V_{\rm end}$}. Note, however, that the
gravitino bound (\ref{ag}) can chop off the lowest part of the above range
by about a couple of orders of magnitude if it is not efficiently diluted by
the curvaton decay.

\begin{figure}
\begin{center}
\leavevmode
\hbox{
\epsfxsize=5in
\epsffile{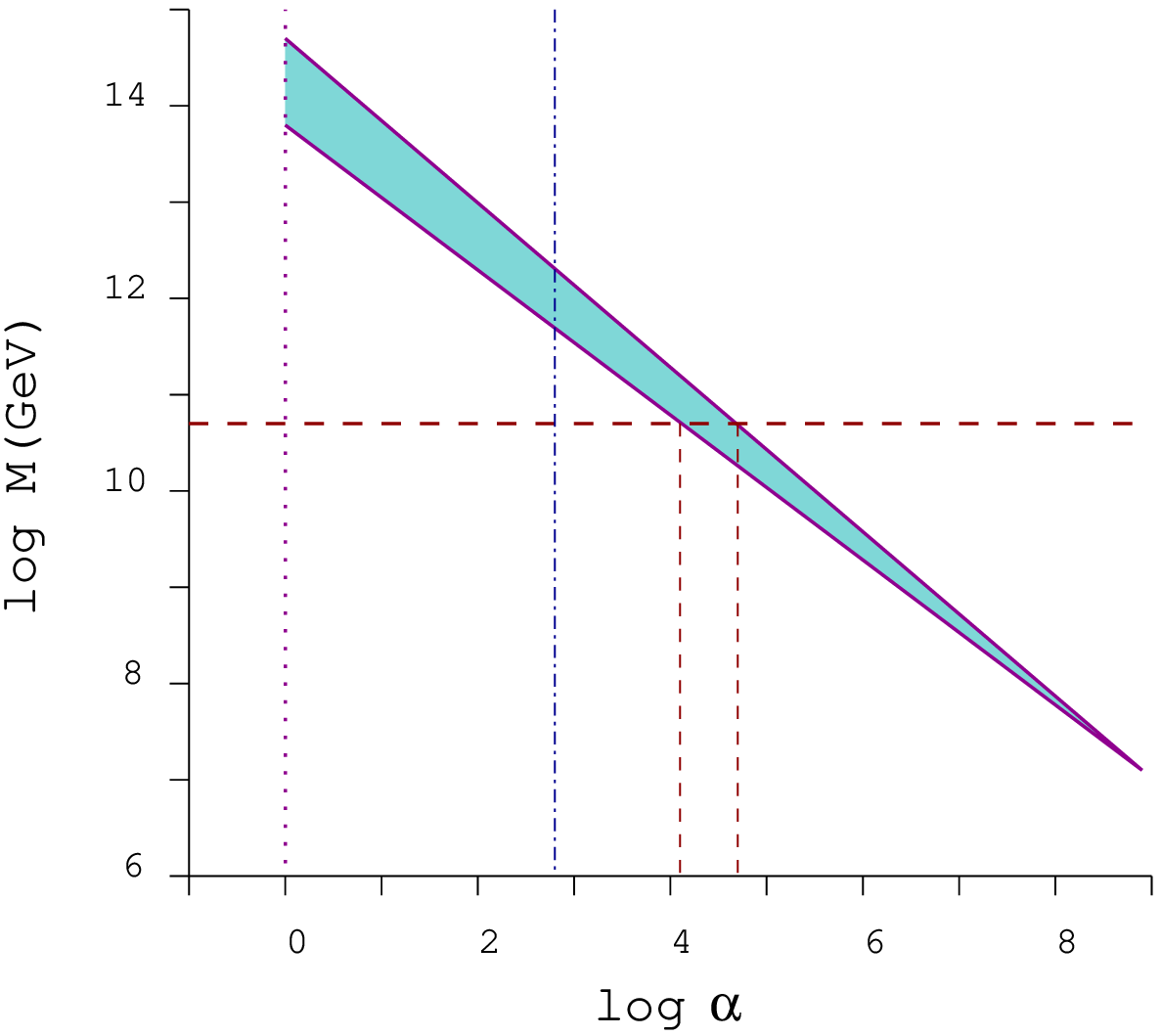}}
\end{center}
\caption{\footnotesize
The shaded region depicts the allowed parameter space for the inflationary
scale $M$. The parameter space is bounded from below by the requirements of
BBN and from above by the {\sc cobe} bound (solid lines). The bounds meet at 
\mbox{$\alpha\simeq 10^{10}$}, which corresponds to 
\mbox{$M_{\sc min}\simeq 10^7$GeV}. The lower bound on $\alpha$ is set by 
the GW~constraint which results in \mbox{$\alpha\geq 1$} (dotted line). The
dashed horizontal line depicts the case of modular quintessential inflation,
for which \mbox{$M=5\times 10^{10}$GeV}. The vertical dashed lines correspond
the the $\alpha$-range for modular quintessential inflation. The vertical 
dashed--dotted line corresponds to the gravitino lower bound on $\alpha$, if 
not diluted by the curvaton's decay.
}
\label{Mplot}
\end{figure}

From (\ref{mmin}) and (\ref{mmax}) we find the following range for the 
inflationary scale for a given $\alpha$

\begin{equation}
6.5\times 10^{13}c^{-3/16}\alpha^{-3/4}{\rm GeV}\leq M\leq
6.0\times 10^{14}c^{-3/14}\alpha^{-6/7}{\rm GeV}
\label{Mrange}
\end{equation}
which is shown in Figure~\ref{Mplot}. We see that entirely uncorrelated 
physics (BBN and coincidence requirements) conspires to allow only a rather 
narrow range for $M$. The range ends up at $\alpha_{\rm max}$, which 
corresponds to the smallest possible value for $M$, which is

\begin{equation}
M_{\sc min}=1.2\times 10^7{\rm GeV}
\label{MMIN}
\end{equation}

The fact that the curvaton hypothesis ameliorates the $\eta$-problem is
related to the value of $q_{\rm min}$. In conventional inflation the 
{\sc cobe} bound is to be saturated and \mbox{$q=q_{\rm min}$}. However,
the spectral index bound (\ref{nscons}), in view of (\ref{ns}), demands
that \mbox{$|\eta|\leq 1/20$}, which, according to (\ref{etaeps}) requires

\begin{equation}
q>20\;b^2
\label{qbound}
\end{equation}
which is impossible to satisfy with $q_{\rm min}$ in the given ranges for
$b,\mu$ and $\alpha$. In contrast, the spectral index constraint 
(\ref{epscons}) in the curvaton case is well satisfied in the allowed 
parameter space. This difference will become apparent in the examples below.

\subsubsection{Examples}

\paragraph{The modular case}
In this case inflation is of the intermediate energy scale which means that
\mbox{$M=5\times 10^{10}$GeV}. %and, therefore \mbox{$\mu=17.70$}. 
Then, using (\ref{Mrange}) one can find the allowed range for $\alpha$:

\begin{equation}
1.4\times 10^4c^{-1/4}\leq\alpha\leq 5.7\times 10^4c^{-1/4}
\end{equation}
which is rather narrow but it is above the gravitino bound (\ref{ag}). Let us 
choose \mbox{$\alpha\equiv 3\times 10^4c^{-1/4}$}. Then, from (\ref{coinc1}) 
we find

\begin{equation}
b=4.56
\end{equation}

Using the above (\ref{qmin}) gives

\begin{equation}
q_{\rm min}=46
\end{equation}
which is quite large but cannot be compared to the requirements of conventional
inflation, which, according to (\ref{qbound}), would demand \mbox{$q\geq 416$}!
Thus, we see that {\em modular quintessential inflation can be realized only 
in the context of the curvaton hypothesis}. This is because, with 
\mbox{$q=q_{\rm min}$}, $|\eta|$ is too large to achieve the required 
almost-scale invariant spectrum of curvature perturbation. Therefore, 
{\em the curvaton hypothesis is necessary to overcome the $\eta$-problem of 
quintessential inflation in conventional cosmology}. Although, strictly 
speaking, the above results have been obtained in the context of the 
toy--model of (\ref{cosh}), we believe that they are generally true for 
models of the type (\ref{modqinf0}) because, as mentioned in Sec.~\ref{kin}, 
the dynamics of $\phi$ are oblivious to the potential during kination and, 
therefore, only the limits of large/small $\phi$, as depicted in 
(\ref{modqinf}), are important.

To obtain an estimate of all the quantities involved in the problem let us
choose \mbox{$q=48$} and \mbox{$c\approx 2$}. Then, from (\ref{scales}) we find

\begin{eqnarray}
m_0 = 10.5\;m_P\approx 2\,M_P & \quad {\rm and} \quad&
m= 0.7\;{\rm TeV}
\end{eqnarray}
which are both rather natural. Using these we also find

\begin{eqnarray}
T_{\rm reh}= 2\times10^6{\rm GeV} 
& \quad {\rm and} \quad &
T_*=100\;{\rm MeV}
\end{eqnarray}
As pointed out earlier, both these values are overestimated by about an order 
of magnitude because of the oversimplified assumption of sudden transition 
from inflation to kination. Still, note that the gravitino constraint
on $T_{\rm reh}$ is well satisfied, as well as the BBN constraint on $T_*$.

From (\ref{fF}) we also find,

\begin{equation}
\phi_F\simeq\left(9.01+\sqrt{q}/16.2\right)M_P%\gsim 11.4\;M_P
\end{equation}
which, for \mbox{$q=48$} gives \mbox{$\phi_F\simeq 9.44\;M_P$}. If modular
quintessential inflation is indeed based on a string model, then the correct 
phenomenology would determine $q$ such that $\phi_F$ is appropriate. The above 
value corresponds to rather large extra dimensions and, therefore, it is not 
clear whether it may be accommodated in a realistic string theory. 

Finally, in view of (\ref{N}), the total number of fast-roll inflationary 
e-foldings is

\begin{equation}
N_{\rm tot}\simeq\frac{1}{F}
\ln\left(\frac{\phi_{\rm end}}{\phi_{\rm in}}\right)
\label{Ntot}
\end{equation}
where \mbox{$\phi_{\rm in}\simeq H_{\rm inf}/2\pi$} because the rolling phase
begins after the inflaton is ``kicked'' away from the origin by its quantum 
fluctuations. Using \mbox{$|\eta|\simeq b^2/q$} and (\ref{fend}) we find 
\mbox{$N_{\rm tot}\approx 100$}. This has to be compared to the number of 
e-foldings that correspond to the horizon at present, which, similarly to 
(\ref{Ndec}), is found to be \cite{jose}

\begin{equation}
N_H=69.15-\ln\alpha
\label{NH}
\end{equation}

Thus, we find \mbox{$N_H\approx 59< N_{\rm tot}$} and the horizon problem is
solved without danger of approaching super-Planckian densities during 
inflation.

\paragraph{\boldmath The case of $M_{\sc min}$}

As another example we consider the case with the smallest possible 
$q_{\rm min}$. From (\ref{qmin}) it is evident that $q_{\rm min}$ decreases
with $\mu$. Therefore, for the smallest $q_{\rm min}$ we need to consider the 
smallest possible value of $M$, which is given by (\ref{MMIN}). This value 
corresponds to $\alpha_{\rm max}$ as given by (\ref{amax}) and also to 
$b_{\rm max}$ as given by (\ref{brange}). Putting all these together 
(\ref{qmin}) gives

\begin{equation}
q_{\sc min}=26
\end{equation}
This should be contrasted with the conventional inflation requirement 
(\ref{qbound}) which demands \mbox{$q>480$}! Thus, again, we see that the 
curvaton hypothesis is necessary to ameliorate the $\eta$-problem.

In order to obtain estimates for the quantities of the problem let us choose 
\mbox{$q=28$} and \mbox{$c\approx 2$}. Then we find

\begin{eqnarray}
m_0 = 5.7\;m_P\approx M_P & \quad {\rm and} \quad&
m= 51\;{\rm keV}
\end{eqnarray}

Using these we also find

\begin{eqnarray}
T_{\rm reh}= 4\;{\rm TeV} 
& \quad {\rm and} \quad &
T_*=10.3
\;{\rm MeV}
\end{eqnarray}
which, again, are both overestimated by an order of magnitude, but satisfy all
the relevant constraints anyway. As before, using (\ref{fF}), we find 
\mbox{$\phi_F\simeq 7.32\;M_P$}. Finally, in a similar manner as above we
find \mbox{$N_{\rm tot}\approx 79$}, which is larger than 
\mbox{$N_H\approx 48$} as required in order to solve the horizon problem.

\section{The case of Brane Cosmology}

\subsection{Brane inflation}

We turn now our attention to the case of brane-cosmology. In this case 
the inflationary dynamics occurs on energy scales higher than the brane tension
(otherwise there would be no difference with the conventional case). Brane 
inflation has been studied in \cite{steep}\cite{roy}. Here we simply cite some 
of the necessary tools to be used in our quintessential inflationary model 
building. 

Above the string tension scale the slow roll parameters are modified and reed

\begin{eqnarray}
\epsilon\equiv\frac{m_P^2}{2}\left(\frac{V'}{V}\right)^2
\frac{1+V/\lambda}{(1+V/2\lambda)^2} & \Rightarrow &
\epsilon\simeq 2\lambda m_P^2\frac{(V')^2}{V^3}\nonumber\\
 & & \label{bsr}\\
\eta\equiv m_P^2\left(\frac{V''}{V}\right)\frac{1}{1+V/2\lambda}
 & \Rightarrow &
\eta\simeq 2\lambda m_P^2\frac{V''}{V^2}
\nonumber
\end{eqnarray}

Similarly the {\sc cobe} constraint (\ref{cobe}) becomes

\begin{equation}
\frac{1}{\sqrt{3}\pi}\frac{V^{3/2}}{m_P^3|V'|}
\left(1+\frac{V}{2\lambda}\right)^{3/2}\simeq
\frac{1}{2\sqrt{6}\pi}\frac{V^3}{\lambda^{3/2}m_P^3|V'|}
\leq 10^{-5}
\label{bcobe}
\end{equation}

Finally, the number of slow-roll e-foldings before the end of inflation is 
related to the value of the scalar field at that time $\phi_N$ by

\begin{equation}
N=\frac{1}{m_P^2}\int_{\phi_N}^{\phi_{\rm end}}\frac{V}{|V'|}
\left(1+\frac{V}{2\lambda}\right)d\phi\simeq
\frac{1}{m_P^2}\int_{\phi_N}^{\phi_{\rm end}}\frac{V^2}{2\lambda|V'|}d\phi
\label{bN}
\end{equation}

\subsection{Exponential quintessential inflation}

\subsubsection{The model}

It can be checked that for models of the form of (\ref{cosh}) or even
steep models such as \mbox{$V(\phi)=M^4[\sinh(\phi/m_0)]^{-q}$} the 
inflationary period already lies in the exponential branch of the potential.
Thus, it is reasonable to avoid complicated models and consider a pure 
exponential potential:

\begin{equation}
V(\phi)=M^4\exp(-b\phi/m_P)
\label{exp}
\end{equation}

The above is well motivated for string moduli due to the considerations of 
Sec.~\ref{modular} (but without the discrete symmetry that forms 
the maximum for $V$). For other motivations of exponential potentials from
Kaluza-Klein, scalar tensor or higher-order gravity theories see, for example,
\cite{exp} and references therein.

For the model (\ref{exp}) the slow roll parameters are

\begin{equation}
\eta=\epsilon=2A\exp(b\phi/m_P)
\label{epseta}
\end{equation}
where \mbox{$A\equiv b^2(\lambda/M^4)$}. Hence we obtain:

\begin{equation}
\phi_{\rm end}=\frac{1}{b}\ln(1/2A)\,m_P
\label{bfend}
\end{equation}
and also

\begin{equation}
V_{\rm end}=2AM^4=2b^2\lambda
\label{bVend}
\end{equation}

Then, using (\ref{bN}) we find

\begin{equation}
\phi_N=-\frac{1}{b}\ln[2A(N+1)]\,m_P
\label{fN}
\end{equation}
and 

\begin{equation}
V(\phi_N)=V_{\rm end}(N+1)
\label{VN}
\end{equation}

\subsubsection{The constraints}

Keeping $A$ a free parameter, we will attempt to constrain the string tension
$\lambda$. Let us begin with the coincidence constraint (\ref{coinc}). 
Defining \mbox{$z\equiv\ln(m_P/\lambda^{1/4})$} and after some algebra we find 

\begin{equation}
z=\frac{56.79-b(1.88-\frac{1}{4}\ln c-\ln\alpha-\frac{5}{3}\ln b)+
\frac{1}{\sqrt{6}}\ln b}{\sqrt{\frac{2}{3}}+b}
\label{zcoinc}
\end{equation}
which diminishes with $b$. Thus, we can define 
\mbox{$z_{\rm min}=z(b_{\rm max})$}. Using \mbox{$b_{\rm max}=2\sqrt{6}$}
we find

\begin{equation}
z_{\rm min}=10.713+\frac{3}{14}\ln c+\frac{6}{7}\ln\alpha
\label{zmin}
\end{equation}

Similarly to the previous section the BBN constraint (\ref{bbn}) can be used
to provide an upper bound to $z$. Indeed, with a bit of algebra we obtain

\begin{equation}
z_{\rm max}=10.706+\frac{3}{16}\ln c+\frac{5}{4}\ln b+\frac{3}{4}\ln\alpha
\label{zmax}
\end{equation}

From the above it is evident that, once more, uncorrelated physics results
in a rather slim parameter space. This parameter space diminishes with 
$\alpha$. Thus, we can find $\alpha_{\rm max}$ by setting
\mbox{$z_{\rm min}=z_{\rm max}$} (or equivalently 
\mbox{$b_{\rm min}=b_{\rm max}$} in (\ref{bbmin}), where now 
\mbox{$Y=\ln(2b^2)$}). We find,

\begin{equation}
\ln\alpha_{\rm max}=18.47-\frac{1}{4}\ln c\Rightarrow
\alpha_{\rm max}\approx 10^8c^{-1/4}
\label{bamax}
\end{equation}

The lower bound on $\alpha$ is set by the GW~constraint (\ref{bGW}), which 
gives, \mbox{$\alpha_{\rm min}\approx 1.5(b/c)^{1/4}$}. Therefore, the 
$\alpha$-range is

\begin{equation}
1\leq\alpha\leq 10^8
\label{barange}
\end{equation}

\begin{figure}
\begin{center}
\leavevmode
\hbox{
\epsfxsize=5in
\epsffile{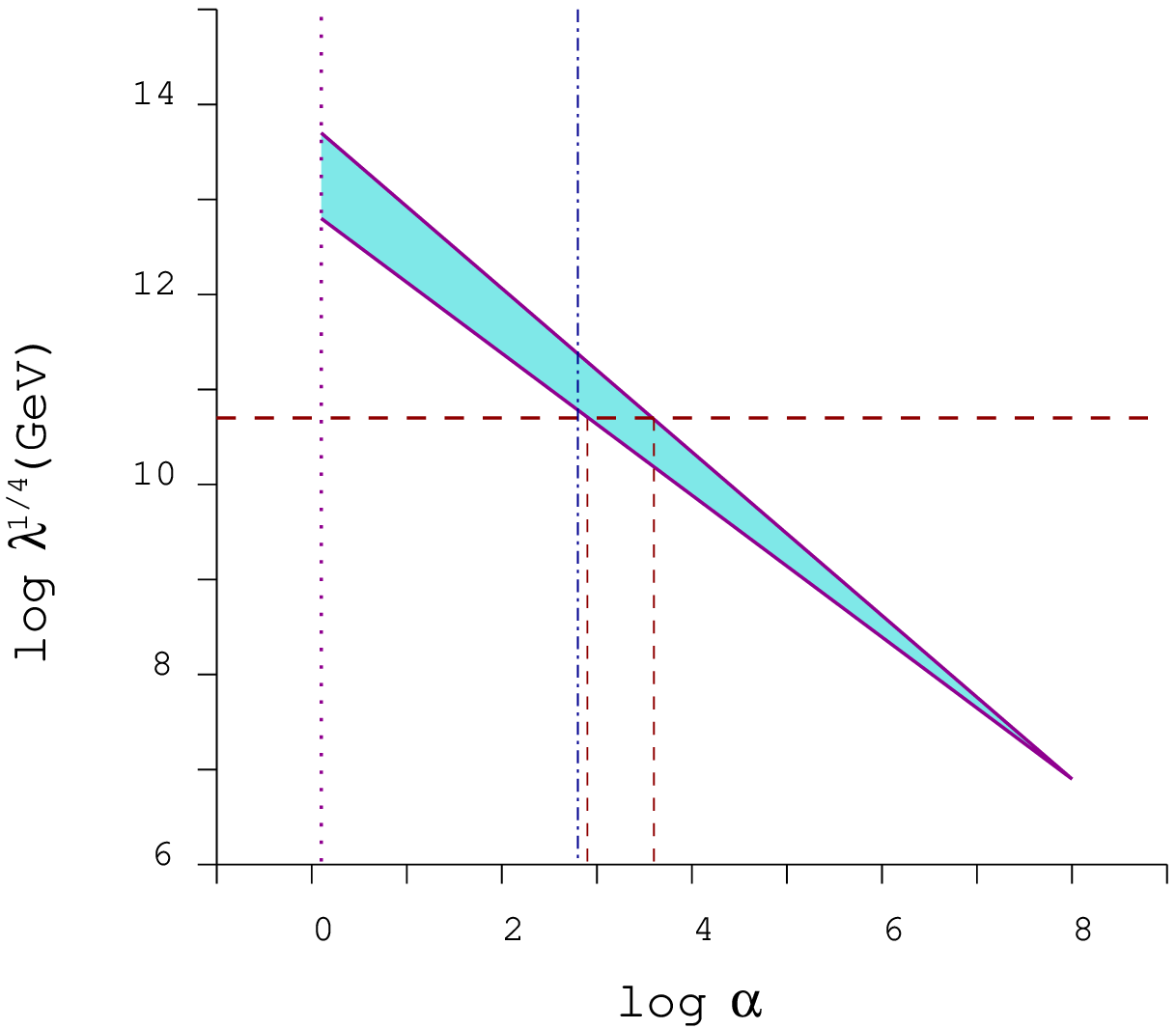}}
\end{center}
\caption{\footnotesize
The shaded region depicts the allowed parameter space for the string
tension $\lambda^{1/4}$ which is linked to the inflationary scale as
\mbox{$V^{1/4}_{\rm end}=\sqrt{b}(2\lambda)^{1/4}$}. 
The parameter space is bounded from below by the requirements of
BBN and from above by the {\sc cobe} bound (solid lines). The bounds meet at 
\mbox{$\alpha\simeq 10^8$}, which corresponds to 
\mbox{$\lambda_{\sc min}^{1/4}\simeq 10^7$GeV}. The lower bound on $\alpha$ is 
set by the GW~constraint which results in \mbox{$\alpha\geq 1$} (dotted line). 
The dashed horizontal line depicts the case where the string tension is of the 
order of the intermediate scale, for which 
\mbox{$\lambda^{1/4}=5\times 10^{10}$GeV}. The vertical dashed lines correspond
the the $\alpha$-range for this case. The vertical 
dashed--dotted line corresponds to the gravitino lower bound on $\alpha$, if 
not diluted by the curvaton's decay.
}
\label{lplot}
\end{figure}

In view of the above the acceptable range for the brane tension, for a given
$\alpha$ is

\begin{equation}
7.5\times 10^{12}c^{-3/16}\alpha^{-3/4}{\rm GeV}\leq \lambda^{1/4}\leq
5.4\times 10^{13}c^{-3/14}\alpha^{-6/7}{\rm GeV}
\label{lrange}
\end{equation}
which is shown at Figure~\ref{lplot}. Note that this range does not differ
much from (\ref{Mrange}). This is so because both are determined by the BBN 
and coincidence constraints on $V_{\rm end}$, which see only the exponential
behaviour of the potential. The difference in the $\alpha$--range, however, is
due to the modified dynamics of brane-cosmology. The parameter space is 
somewhat reduced in size because of the negative effect of overdamping during 
kination.

The smallest possible $\lambda$ 
corresponds to $\alpha_{\rm max}$. Using (\ref{bamax}) we find,

\begin{equation}
\lambda_{\sc min}^{1/4}=7.2\times 10^6{\rm GeV}
\end{equation}
which, in view of (\ref{M5}), corresponds to \mbox{$M_5\approx 10^{11}$GeV}.
Now, the {\sc cobe} bound, as given by (\ref{bcobe}), becomes

\begin{equation}
z\geq 5.08+\frac{3}{2}\ln b+\ln(N_{\rm dec}+1)
\label{zcobe}
\end{equation}
where $N_{\rm dec}$ is again given by (\ref{Ndec}). It can be shown that
the above does not change drastically over the $\alpha$-range 
(when increasing $\alpha$ the mild growth of the $b_{\rm min}$ is counteracted 
by the decrease of $N_{\rm dec}$) and corresponds to an overall bound

\begin{equation}
\lambda^{1/4}\leq 2.6\times 10^{13}{\rm GeV}
\label{cobebound}
\end{equation}
which is satisfied over all the range (\ref{lrange}). This bound is challenged
and may be saturated only for \mbox{$\alpha\approx 1$}, which, however, is in
danger to violate the GW~constraint (and will certainly violate the gravitino 
constraint (\ref{ag}) if it is applicable). Thus, we see that {\em without the 
curvaton hypothesis one can hardly secure any parameter space for successful 
quintessential inflation}. Moreover, note that, in the context of the curvaton 
hypothesis, the GW and the gravitino constraints are somewhat relaxed by the 
entropy production due to the curvaton decay.

It can be checked that within the above parameter space a number of other 
constraints that apply to the system are well satisfied. In particular, one
does not violate the prompt reheating constraint 
\mbox{$\rho_{\rm reh}\leq V_{\rm end}$}. Also there is an absolute upper bound 
on $\lambda$ coming from \mbox{$\lambda^{1/4}\leq M_5$}, which, in view of
(\ref{M5}) is recast as

\begin{equation}
\lambda^{1/4}\leq\frac{8\pi}{\sqrt{6}}\,m_P\approx 2\,M_P
\end{equation}
which is obviously satisfied. Another relevant bound is 
\mbox{$V(\phi_{\rm in})<M_5^4$}. Using (\ref{VN}) and (\ref{bVend}) we see
that this bound corresponds to 

\begin{equation}
N_{\rm tot}<\frac{M_5^4}{2b^2\lambda}-1\equiv N_{\rm max}
\end{equation}
Using (\ref{M5}) and (\ref{NH}) it can be shown that 
\mbox{$N_H\ll N_{\rm max}$}, throughout all the above parameter space and,
therefore, the horizon problem is solved without problems.

As far as the spectral index is concerned it can be shown that the 
observational requirement (\ref{nscons}) is not challenged in both 
conventional inflation and, of course, in the context of the curvaton 
hypothesis. Indeed, in conventional inflation we have 
\mbox{$n_s-1=-\frac{4}{N+1}$}, which means that (\ref{nscons}) sets the bound
\mbox{$N_{\rm dec}\geq 39$}. Using (\ref{Ndec}) this bound translates into
\mbox{$\alpha\leq 10^{12}$}, which is true for all the parameter space of 
interest (c.f. (\ref{barange})). Similarly, for the curvaton case and ignoring
$\eta_{\sigma\sigma}$ we obtain the bound \mbox{$\alpha\leq 10^{21}$}, which is
well beyond challenge. Thus, we see that, in the case of brane quintessential
inflation the benefits of the curvaton hypothesis are related more to the
possible reduction of the inflationary scale (allowed from the {\sc cobe} 
bound) than to the $\eta$-problem itself. This is because steep-inflation does
help reducing $|\eta|$ as long as the inflationary scale can be lowered to
counteract the effect of overdamping which reduces the duration of kination. 
By relaxing the {\sc cobe} constraint into an upper bound, the curvaton 
hypothesis enables us to do just that.

Finally, it should be stressed that $M$ can be anything as long as 
\mbox{$M\leq M_5$}, which results into the constraint:

\begin{equation}
A\geq\left(\frac{\sqrt{6}}{8\pi}\right)^{4/3}b^2
\left(\frac{\lambda^{1/4}}{m_P}\right)^{4/3}
\label{Abound}
\end{equation}

\subsubsection{Example}

Let us consider again the intermediate scale, 
\mbox{$\lambda^{1/4}=5\times 10^{10}$GeV}.
In this case (\ref{M5}) gives \mbox{$M_5=4\times 10^{13}$GeV}.
Then, from (\ref{lrange}) we find the following range for $\alpha$:

\begin{equation}
0.8\times 10^3c^{-1/4}\leq\alpha\leq 3.5\times 10^3c^{-1/4}
\end{equation}
which, again is above the gravitino bound (\ref{ag}).
Let us choose \mbox{$\alpha=2\times 10^3c^{-1/4}$}. Then, using 
(\ref{zcoinc}), we obtain

\begin{equation}
b=4.54
\end{equation}

Using this and taking \mbox{$c\approx 2$} we find:

\begin{eqnarray}
T_{\rm reh}=\frac{\alpha b^2}{\sqrt{6\pi}}\frac{\sqrt{\lambda}}{m_P}
 & \Rightarrow & T_{\rm reh}=8\times 10^6{\rm GeV}\\
 & & \nonumber\\
T_*=\frac{2\alpha^3b^5}{(12\pi)^2}\sqrt{\frac{2g_{\rm reh}}{5}}
\left(\frac{g_{\rm reh}}{g_*}\right)^{1/4}\frac{\lambda}{m_P^3}
 & \Rightarrow & T_*=110\;{\rm MeV}
\end{eqnarray}
which are, again, overestimated by an order of magnitude, but still satisfy 
all the constraints, such as the gravitino bound and the BBN constraint. Also,
note that $T_{\rm reh}$ is well below the so--called normalcy temperature 
\mbox{$T_c\simeq\lambda^{1/4}$} \cite{normalcy}, above which Kaluza-Klein 
excitations on the brane may radiate energy into the bulk and possibly 
reinstate the dark radiation term in (\ref{friedbrane}). 

Now, the $A$-bound (\ref{Abound}) reeds

\begin{equation}
A\geq 5.2\times 10^{-11}
\label{Afin}
\end{equation}
which, when saturated, results in \mbox{$M=M_5$}. Using (\ref{bfF}) we find
\mbox{$\phi_F=\phi_{\rm end}+9.20\;M_P$}, which, in view of (\ref{bfend}) and 
(\ref{Afin}) gives \mbox{$\phi_F\leq 10.21\;M_P$}. A preferred value  of 
$\phi_F$ may be achieved by adjusting $A$, or, equivalently $M$. For example,
for \mbox{$M=1$ TeV} we have \mbox{$A=1.3\times 10^{32}$} and 
\mbox{$\phi_F\simeq 5.92\;M_P$}.

\section{Conclusions}

We have investigated the $\eta$-problem of quintessential inflation 
model-building. In the context of a potential with an exponential 
quintessential tail we have shown that brane cosmology inhibits the 
period of kination due to the extra friction on the roll-down of the scalar 
field. This counteracts the beneficial effects of steep inflation towards 
overcoming the $\eta$-problem. Hence, we pursued a different approach and 
considered quintessential inflation in the context of the curvaton hypothesis. 
We showed that the latter substantially ameliorates the $\eta$-problem
in both the cases of conventional and brane cosmology. To demonstrate this
we have studied a toy model of what we called modular quintessential inflation
in the case of conventional cosmology and the pure exponential potential in
the case of brane-cosmology. In both cases we have shown that the available
parameter space for the inflationary scale is not large and it is strongly 
correlated with the reheating efficiency $\alpha$. Indeed, for a given 
$V_{\rm end}$, we have shown that there is only a small window for $\alpha$,
where successful quintessential inflation is possible. This may seem like a 
fine-tuning problem. However, it simply reflects the necessary tuning for 
successful coincidence. The required values for $\alpha$ are not unreasonable 
and we should point out that there is nothing special about the present time.
Any value of $\alpha$ would cause some brief acceleration period in the late 
Universe. We just happen to live in this period. These tuning considerations 
are even more relaxed if one considers the possibility of multiple unfreezings 
and refreezings of the scalar field, as discussed at the end of 
Sec.~\ref{coincbbn}. 

In this paper we have considered the intriguing possibility that the scalar 
field of quintessential inflation (called the `cosmon' by some authors) is
a modulus field, possibly associated with the volume of the extra dimensions,
such as the geometrical $T$-moduli of weakly coupled heterotic string 
theory. The modulus is taken to roll down and away from the origin, where it 
could have been placed by temperature corrections to its potential during a 
period of thermalization preexisting inflation, if the origin is a point of 
enhanced symmetry. In this scenario the inflationary expansion begins with a 
period of thermal inflation followed by fast-roll inflation, as described in 
\cite{mine} for modular thermal inflation. In contrast to \cite{mine} though, 
we have supposed that the K\"{a}hler corrections introduce an exponential 
slope to the potential over distances comparable to $m_P$ in field space. 
Thus, the VEV of the modulus is displaced at infinity, while the modulus is 
stabilized dynamically by being frozen during the later history of the 
Universe at a non-zero potential density, causing the 
present accelerated expansion. This way it may be natural to avoid the 
excessive supergravity corrections that would otherwise increase the present 
mass of quintessence to unacceptable values. However, it remains to be seen 
whether this scenario is possible in the context of a realistic string theory.

Turning to brane cosmology we have focused in the much investigated 
pure exponential potential, which may also be motivated by string theory 
considerations. In this case there is no preferred starting point for the
roll down of the field as long as the inflationary energy scale is kept below 
the fundamental scale of the theory. We have seen that the parameter space
for successful quintessential inflation is somewhat reduced by the 
negative overdamping effect of brane-cosmology on kination.

Finally, we have studied the effects of gravitational wave generation on 
quintessential inflationary model-building. We have shown that gravitational 
waves will not destabilize BBN if the reheating efficiency is 
\mbox{$\alpha>1$}, which may require some tiny, but non-zero coupling of the 
inflaton with other fields. In the context of the curvaton hypothesis, 
however, the gravitational wave constraint is ameliorated by the dilution 
effect of the entropy production due to the curvaton's decay.
This may lower the bound on $\alpha$ below \mbox{$\alpha\sim 0.1$}, which will
render gravitational reheating (and a truly sterile inflaton) acceptable.
However, a larger $\alpha$ may be necessary in order to avoid gravitino 
overproduction: \mbox{$\alpha\gsim 10^2$}. Note, here, that tiny couplings 
between the inflaton and the SM fields may have beneficiary side effects, such 
as baryogenesis \cite{baryon}.

All in all we have shown that the liberating effect of the curvaton hypothesis 
enables quintessential inflation to overcome its $\eta$-problem and 
enlarges the parameter space for successful model-building. 
Appealing candidates for the quintessential inflaton (or cosmon) may be 
string--moduli fields.

\bigskip

\noindent
{\Large\bf Acknowledgments}

\nopagebreak[4]

\medskip

\nopagebreak[4]

\noindent
I would like to thank D.H.~Lyth and J.E.~Lidsey for discussions. This work was
supported by the E.U.~network program: HPRN-CT00-00152.

\end{document}